\documentclass[showpacs,preprintnumbers,amsmath,amssymb,amsthm]{elsarticle}

\usepackage{amssymb,amsmath,amsthm}
\usepackage{graphicx}
\DeclareGraphicsExtensions{.eps,.bmp}
\usepackage{dcolumn}
\usepackage{bm}
\usepackage{txfonts}
\usepackage{subfigure}
\usepackage[dvips]{psfrag}
\usepackage{float}
\usepackage{color}
\usepackage[usenames,dvipsnames,svgnames,table]{xcolor}
\biboptions{sort&compress}
\DeclareMathAlphabet{\mathcal}{OMS}{cmsy}{m}{n}

\def\Cu{{\rm Cu}}\def\d{{\rm d}}

\begin{document}
\graphicspath{{../FIGURAS6/}}

\title{Waving transport and propulsion in a generalized Newtonian fluid}

\author[vel]{J. Rodrigo V\'elez-Cordero}
\address[vel]{Instituto de Investigaciones en Materiales, Universidad Nacional Aut\'onoma de M\'exico, Apdo. Postal 70-360
M\'exico D.F. 04510.}
\author[rvt]{Eric Lauga}
\address[rvt]{Department of Applied Mathematics and Theoretical Physics, University of Cambridge, Centre for
Mathematical Sciences, Wilberforce Road, Cambridge, CB3 0WA, UK.}

\date{\today}

\begin{abstract}
Cilia and flagella are hair-like appendages that protrude from the surface of a  variety of  eukaryotic cells and deform in a
wavelike fashion to transport fluids and propel cells. Motivated by the ubiquity of non-Newtonian fluids in biology, we
address mathematically the role of shear-dependent viscosities on both the waving flagellar locomotion and ciliary transport
by metachronal waves. Using a two-dimensional waving sheet as model for the kinematics of a flagellum or an array of cilia,
and allowing for both normal and tangential deformation of the sheet, we calculate the flow field induced by a
small-amplitude deformation of the sheet in a generalized Newtonian Carreau fluid up to order four in the dimensionless
waving amplitude. The net flow induced at far from the sheet can be interpreted either as a net pumping flow or, in the frame
moving with the sheet,  as a swimming velocity. At leading order (square in the waving amplitude), the net flow induced by
the waving sheet and the rate of viscous dissipation is the same as the Newtonian case, but  is different  at the next
nontrivial order (four in the waving amplitude). If the sheet deforms both in the  directions perpendicular and parallel to
the wave progression, the shear-dependence of the viscosity leads to a nonzero flow induced  in the far field while if the
sheet is inextensible, the non-Newtonian influence is exactly zero. Shear-thinning and shear-thickening fluids are seen to
always induce opposite effects. When the fluid is shear-thinning, the rate of working of the sheet against the fluid is
always smaller than in the Newtonian fluid, and the largest gain is obtained for antiplectic metachronal waves. Considering a
variety of deformation kinematics for the sheet, we further show that in all cases transport by the sheet is more efficiency
in a shear-thinning fluid, and in most cases the transport speed in the fluid is also increased. Comparing the order of
magnitude of the shear-thinning contributions  with past work on elastic effects as well as the magnitude of the Newtonian
contributions, our theoretical results, which beyond the Carreau model are valid for a wide class of generalized Newtonian
fluids, suggest that the impact of shear-dependent viscosities on transport could play a major biological role.

\end{abstract}

\maketitle

\section{Introduction}

The editors of the recently-created journal \emph{Cilia} called their inaugural article ``Cilia - the prodigal organelle''
\cite{Beales2012}. This is perhaps an appropriate denotation for a biological appendage found on the surface of a variety of
eukaryotic cells,   with  a fascinating and still not fully understood  relationship between internal structure and
biological function \cite{Lindemann2010}. Cilia  not only have the capacity  to transport surrounding fluids by means of
periodic movements but they also play important  sensory roles \cite{Bloodgood2010,Beales2012}.  Remarkably, their basic
morphology and internal structure (the so-called axoneme), which is the same as in all eukaryotic flagella, has been
conserved throughout evolution \cite{Ibanez2003}.

The active dynamics of cilia plays a crucial role in a  major aspect of biology, namely fluid transport
\cite{Brennen1977,Sleigh1988,Ibanez2003}.  The motion of biological fluids induced by the collective beating of an array of
cilia  is the third major form of mechanical fluid transport done by the inner organs of  superior animals, ranking only
behind  blood pumping and peristaltic motion. In humans, the flow produced by the deformation of cilia is involved in the
transport of several biological fluids, including the removal of tracheobronchial mucus in the respiratory track
\cite{Brennen1977,Sleigh1988,Ibanez2003}, the transport of ovulatory mucus and the ovum in the oviduct of the female
reproductive track \cite{Ibanez2003,Fauci2006}, and the motion of epididymal fluid in the efferent ducts of the male
reproductive track \cite{Lardner1972}. Failure of the transport functionality of cilia  can lead  to serious illness of the
respiratory system \cite{Ibanez2003}. In addition, it is  believed that cilia found in specialized brain cells (ependymal
cilia) are  involved in the transport of cerebrospinal fluid at small scales \cite{OCallaghan2012}.  When cilia are anchored
on a free-swimming cell, such as in many protozoa including the oft-studied {\it Paramecium}, their collective deformation
and  active transport of the surrounding fluid  leads to locomotion of the cell, typically at very low Reynolds number
\cite{Brennen1977,Lauga2009}.

The study of  fluid transport  by cilia,  and more generally by eukaryotic flagella,  requires an understanding of  their
beating patterns as well as the physical properties of the surrounding fluid. Many biological fluids in the examples above
display strong non-Newtonian characteristics, including  nonzero relaxation times and shear-dependent viscosities
\cite{Lai2009}. Past work has addressed theoretically \cite{Lauga2007,FuPowersWolgemuth2007,Fu2009,lauga_life,teran2010,lailai-pre,laipof1} and
experimentally \cite{Shen2011,Liu2011,dasgupta13} the effect of fluid viscoelasticity on transport and locomotion, but  the impact of
shear-dependance material functions on the performance of cilia and flagella has yet to be fully quantified \cite{Smith2012}.

In the present paper we  mathematically address the role of shear-dependent viscosity  on the propulsion performance of a
two-dimensional undulating surface whose boundary conditions model the beating strokes of a cilia array or of a waving
flagellum. The calculation is an extension of the classical results  by Taylor  \cite{Taylor1951,Childress1981}  to the
non-Newtonian case.  In the next section we describe the physics of the cilia beating patterns, distinguish individual
vs.~collective motion, and  introduce the mathematical models that have been proposed to quantify cilia dynamics both in
Newtonian and  non-Newtonian fluids.

\section{Background}

\subsection{Kinematics of an individual cilium}

\begin{figure}[t]
\centering
\includegraphics[scale=0.4]{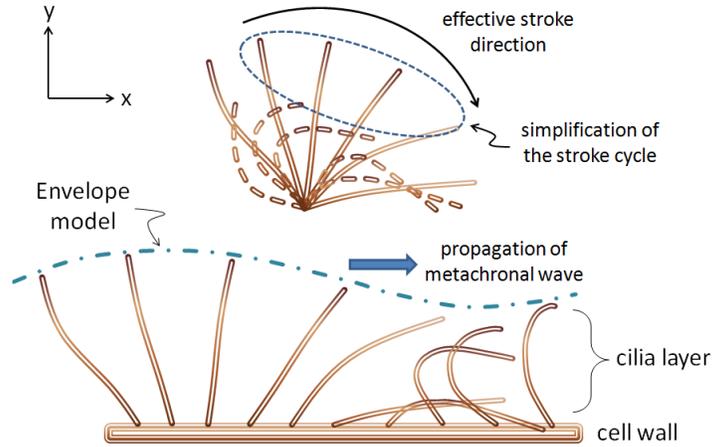}
\caption{Schematic representation of the  stroke of an individual cilium and the envelope model. Top:  effective (solid) and recovery (dashed) stroke of a cilium. The ellipse (thin dashed) represents the stroke cycle as modeled by the envelope model; in this figure the direction of the dynamics along the ellipse is clockwise, but it could be reversed  (see text). Bottom: Representation of the envelope model covering the cilia layer and the propagation of the metachronal wave.} \label{cilia}
\end{figure}

A variety of beating patterns is displayed by cells employing cilia or flagella, depending on the surrounding geometry, the purpose of the beating (locomotion, ingestion, mixing...) as well as genetic factors and biochemical regulation \cite{Brennen1977,Verdugo1982,Takahashi2002,Ibanez2003,Lindemann2010}. A detailed description of the  stroke of cilia has
been offered in the past  \cite{Brennen1977,Sleigh1988,Childress1981,Brennen1974} and we summarize the main features here. In general, the beating pattern of an individual cilium displays a two-stroke effective-recovery motion, as illustrated in Fig.~\ref{cilia} (top). During the effective stroke, the cilium  extends into the fluid  dragging the maximum volume of fluid forward. In contrast, the recovery (or backward) stroke is executed by bending the cilium towards itself and the nearest boundary,  minimizing thereby the drag on the surrounding  fluid in the opposite direction.  It is this asymmetry in the beating pattern that induces  a net fluid flow in the direction of the effective stroke.  In some cells,  the cilium not only bends towards the wall during the recovery stroke but also tilts with respect to the vertical plane described by the effective stroke, hence displaying a three-dimensional pattern \cite{Sanderson1981,Sleigh1988}. Cilia in the respiratory
track seem, however, to remain two-dimensional \cite{Chilvers2000}.

\subsection{Collective cilia motion and metachronal waves}

If we now turn our attention to the collective motion of cilia array, the cilia are observed to beat in an organized manner such that an undulating surface will appear to form on top of the layer of cilia, and to deform in a wave-like fashion. These waves, known as metachronal waves, are due to a small phase lag  between neighboring cilia, and are akin to  waves created in sports stadiums by waving spectators. Many biological studies have been devoted to this collective effect \cite{Murakami1975,Takahashi2002}.  In Table \ref{tableone} we reproduce some typical values of the beating parameters measured  experimentally for  cilia movement in the respiratory track (tracheobronchial cilia). The cilia length and the
distance between cilia at their base are denoted by $L$ and $d$, while $f=\omega/2\pi$ is the beat frequency and $\lambda=2\pi/k$  is  the metachronal wavelength.

Different types of metachronal waves can be  classified according to the relationship between their dynamics and that of the effective stroke of the constituting cilia \cite{Brennen1977}.  When the propagative direction of the metachronal wave is the same as the direction of the effective stroke, the beat coordination is called symplectic. If instead  both directions oppose each other then the coordination is termed  antiplectic  \cite{Childress1981}.  Other types of metachronal waves that requires certain three-dimensional coordination have also been identified \cite{Sleigh1988,Blake1971,Brennen1974,Sanderson1981}. Although there is no yet agreed-upon explanation for the origin of
these different types of coordination, it seems that patterns close to the antiplectic metachrony (including tracheobronchial
cilia) are more widely used than the symplectic wave \cite{Knight1954} possibly because the first one is characterized by the
separation of consecutive cilia during the effective stoke, allowing them to propel more fluid volume and thus increasing the
stoke efficiency \cite{Blake1972}.

\begin{table}[t]
  \centering
  \scalebox{0.9}{\begin{tabular}{c|c|c}
 \quad\quad\quad\quad \quad\quad Parameter\quad\quad\quad\quad\quad\quad &  \quad\quad\quad\quad\quad\quad & \quad Reference\quad \\ \hline
  Cilium length ($L$) & 6 $\mu$m & \cite{Ross1974}\\
  Beating frequency ($f=\omega/2\pi$) & 20 Hz & \cite{OCallaghan2012}\\
  Distance between cilia ($d$) &  0.4 $\mu$m  & \cite{Sleigh1988}\\
  Metachronal wavelength ($\lambda=2\pi/k$) & 30 $\mu$m & \cite{Ross1974}\\   \end{tabular}}
  \caption{Beating parameters for metachronal waves of tracheobronchial cilia: cilia length ($L$),  beating frequency ($f$), distance between cilia ($d$), and metachronal wavelength ($\lambda$). These numbers are found in the references indicated in the last column}
  \label{tableone}
\end{table}

\subsection{Modeling cilia arrays}

Two different approaches have been proposed to model the periodic motion of cilia, namely the sub-layer  and the envelope
models \cite{Childress1981}. In sub-layer modeling,  one keeps  the basic ingredients of the cilia beating cycle by
parametrizing  each cilium shape along its stroke period.  This approach allows  to compute the forces and bending moments
induced by each cilium to the surrounding fluid and at the same time obtain the mean flow produced above the cilia layer,
although the hydrodynamic description can be tedious and usually has to be done numerically
\cite{Blake1972,Blake1975,Lauga2009}.

In contrast, the envelope modeling approach  takes advantage of the formation of metachronal waves above the cilia layer. The
motion of the cilia array is simplified as an undulating surface that covers the cilia layer  (Fig.~\ref{cilia}, bottom),
ignoring the details of the sub-layer dynamics.  In this case, the metachronal waves are prescribed  by setting appropriate
boundary conditions on the material points of the envelope. In the case where the boundary conditions are nearly
inextensible, the dynamics of the flow produced by the cilia array becomes similar to the undulating motion of a
two-dimensional flagellum \cite{Brennen1977}.  The  simplification allowed by the envelope model has been useful in
comparing, even quantitatively, the swimming velocities obtained theoretically with those observed in water for a number of
microorganisms \cite{Brennen1974,Blake1971b}. Furthermore,  this model, which is originally due to Taylor \cite{Taylor1951},
is amenable to perturbation analysis and allows  to incorporate certain non-Newtonian effects in a systematic fashion
\cite{Lauga2007,lauga_life}.

To address the limit of  validity of the envelope model, Brennen \cite{Brennen1974}  derived the set of conditions  to be
satisfied in order for the envelope model to be approximately valid, which are $\nu/({\omega}L^{2})\gg{kd}$ and $kd<1$,
where $\nu$ is the kinematic viscosity of the fluid. With the numbers from Table \ref{tableone}, we see  that the envelope
model is valid for the case of the cilia array  in the respiratory track when the liquid has a viscosity similar to that of
water (and $\nu$ will be actually higher if the fluid is a polymeric solution, hence the model is expected to always be
valid).

\subsection{Cilia in complex fluids}

Because of the importance of mucus in human health, numerous  theoretical papers have attempted  to elucidate  the mechanisms
of tracheobronchial mucus transport using both modeling approaches outlined above
\cite{Barton1967,Ross1974,Blake1975,Fulford1986}. Mucus is a typical example of a non-Newtonian fluid, displaying both
elastic behavior and stress relaxation as well as a shear-dependent viscosity \cite{Lai2009}.

In order to incorporate a variation of the viscosity in the transport problem, a two-fluid model has been proposed
consisting of  a low-viscosity fluid located at the cilia layer (peri-ciliary fluid) with  a second high viscosity fluid
laying above the first one \cite{Sleigh1988,Blake1975,Fulford1986,Rogers1997}. This hypothesis of a two-fluid layer,
supported by experimental observations of tissue samples, allows to obtain values of the mucus flow similar to those measured
in experiments \cite{Sleigh1988,Sanderson1981}.

An alternative modeling approach proposed to use  a fluid with a viscosity whose value changes linearly with the distance
from the cell wall  \cite{Barton1967}. Such approximation, however, is not a  formal mechanical model that rigorously takes into account the rheological properties of the fluid. Later work considered formally the  elastic properties of mucus by introducing the convective Maxwell equation in the theoretical formulation  \cite{Ross1974}. The values of the mean velocity
of the fluid in this case were found to be much smaller than the experimental ones, indicating that the shear-thinning properties of mucus also need to be considered.  Elastic effects were  the focus on a number of studies on flagellar locomotion, both theoretically  \cite{Lauga2007,FuPowersWolgemuth2007,Fu2009,lauga_life,teran2010,lailai-pre,laipof1}  and experimentally \cite{Shen2011,Liu2011,dasgupta13}, and the consensus so far seems to be that for small-amplitude motion elastic effects do always hinder locomotion whereas it can potentially be enhanced  for large amplitude motion corresponding to order-one Weissenberg, or Deborah,  numbers.

For mucus transport by cilia, it is in fact believed that it is a combination of shear-thinning, which can reduce the fluid
flow resistance at the cilia layer, and  elastic effects, which maintain a semi-rigid surface at the upper mucus layer, which
allow  for the optimal transport of particles in the respiratory track
\cite{Sleigh1988,Sanderson1981,Barton1967,Rogers1997,Sade1970}. A constitutive equation that takes into account a
shear-dependent viscosity was formally used within the  lubrication approximation  in the context of gastropod
\cite{Chan2005,Lauga2006}  and flagellar locomotion \cite{Balmforth2010}. These models consider the locomotion due to a
tangential or normal (or  both) deformation of a sheet above a thin fluid layer with shear-dependent viscosity. In the case
of  tangential deformation of the sheet, the average locomotion speed is seen to go down for a shear-thinning fluid whereas
in the case where normal deformation are also included the speed increases.  This indicates that the impact of
shear-dependent material function on transport depends on the particular beating pattern, something our model also shows.
Recent studies on the effect of shear-dependent viscosity on waving propulsion showed that locomotion is unaffected by
shear-thinning for nematodes (experiments, \cite{Shen2011}), increases in a two-dimensional model flagellum with an increase
in flagellar amplitude for shear-thinning fluids (computations, \cite{Smith2012}), while decreases in the case of a
two-dimensional swimming-sheet-like swimmer model in shear-thinning elastic fluids (experiments, \cite{dasgupta13}).

Other models focused on the presence of yield stresses and the heterogeneity in the surrounding environment \cite{Chan2005,Leshansky2009,Balmforth2010}. The presence of a yield stress in the limit of small amplitude oscillations hinders the mean propulsion or transport velocity whereas the inclusion of obstacles mimicking an heterogeneous polymeric solution  enhances propulsion.

\subsection{Outline}

In this paper we solve for the envelope model in a generalized Newtonian fluid using a domain perturbation expansion. The
fluid is the Carreau model for generalized Newtonian fluids where the viscosity is an instantaneous nonlinear function of the
local shear rate. We solve  for the flow velocity and rate of work in powers of the amplitude of the sheet deformation. In
particular, we compute analytically the influence of non-Newtonian stresses on the fluid velocity induced in the far field:
that velocity can be interpreted either as a transport velocity in the context of fluid pumping or as a propulsive  speed in
the context of locomotion.  In section \ref{formulation} we present the formulation of the envelope model and the dynamic
equations that govern the fluid flow. Section \ref{asymptotic} details the asymptotic calculations followed by  section
\ref{analysis} which is devoted to an analysis of our mathematical results and the illustration of the importance of
non-Newtonian effects. Finally in section \ref{discussion} we discuss the impact of our results in the context of biological
locomotion and transport and conclude the paper. In \ref{appendix} the applicability of our results to a wider class of
generalized Newtonian fluid models is examined.

\section{Formulation}
\label{formulation}
\subsection{The envelope model}

In the envelope model approach, the cilia tips, attached to a stationary base, display a combination of normal and tangential
periodic motions. Depending on the phase shift $\phi$ between these two motions and their relative amplitude ($-\pi \leq \phi
\leq \pi$), in general the cilia tips will describe an elliptic shape in the $xy$-plane over one period of oscillation
(Fig.~\ref{cilia}).  We assume that the cilia tips are connected by a continuous waving sheet, and the variation of the
distance between adjacent cilia is therefore idealized as a stretching or compression of the surface. The positions of
material points ($x_{m},y_{m}$) on the sheet are given by
\begin{equation}\label{materialpoints1}
     x_{m}=x+Aa\cos{(kx-\omega{t}-\phi)},  \quad y_{m}=Ab\sin{(kx-\omega{t})},
\end{equation}
where $2Aa$ and $2Ab$ are the maximum displacements of the material points in the $x$ and $y$ directions respectively. The
wave amplitude is thus $A$ (dimensions of length) while the relative values of the dimensionless parameters $a>0$ and $b>0$
reflect the ratio of the tangential and normal motions.  The propagation of the metachronal wave in eq.~\ref{materialpoints1}
occurs in the  positive $x$ direction (Fig.~\ref{cilia}) and the material points of the sheet can describe a cyclic motion
either in a clockwise direction (symplectic), or a counterclockwise direction (antiplectic).

We nondimensionalize lengths by $k^{-1}$ and times by $\omega^{-1}$ to get the nondimensionalized form of
eq.~\ref{materialpoints1} as
\begin{equation}\label{materialpoints2}
\begin{matrix}
     x_{m}=x+\epsilon{a}\cos{(x-t-\phi)}, & y_{m}=\epsilon{b}\sin{(x-t)},
\end{matrix}
\end{equation}
where we have defined $\epsilon=Ak=2\pi{A}/\lambda$.  Assuming that the cilia array propagates metachronal waves with a
single characteristic amplitude much smaller than its wavelength, we can use a standard domain perturbation method to solve
for the pumping velocity in the limit $\epsilon\ll1$.  In the remaining text we will use  for convenience $z=x-t$ and
$\xi=x-t-\phi$. Using the formula $\cos{(z-\phi)}=\cos{z}\cos{\phi}+\sin{z}\sin{\phi}$, we will  also define
$\beta=a\cos{\phi}$, $\gamma=a\sin{\phi}$, and $a^2=\beta^2+\gamma^2$.

\subsection{Governing equations of the fluid flow and boundary conditions}

Neglecting any inertial effects in the fluid, mechanical equilibrium is simply given by $\nabla\cdot\bm{\sigma}={\bf 0}$
where $\bm{\sigma}$ is the total  stress tensor. Denoting
 $p$ the pressure and $\bm{\tau}$ the stress tensor component due to fluid deformation (deviatoric) we have
\begin{equation}
\label{lowReequation} \nabla{p}=\nabla\cdot\bm{\tau},
\end{equation}
which is associated with the  incompressibility condition, $\nabla\cdot\bm{u}=0$, where $\bm{u}$ denotes the velocity field.

In our model, the  fluid problem is two-dimensional, and we can therefore reduce the number of scalars used in the
formulation by introducing a streamfunction, $\psi(x,y,t)$. The  incompressibility condition is satisfied everywhere in the
fluid provided we use
\begin{equation}\label{incompressibility}
     u=\frac{\partial{\psi}}{\partial{y}}, \quad  v=-\frac{\partial{\psi}}{\partial{x}}\cdot
\end{equation}
Using eq.~\ref{incompressibility} and the time derivatives of the material position of the undulating sheet,
$u=\partial{x_{m}}/\partial{t}$ and $v=\partial{y_{m}}/\partial{t}$, we can identify the velocity components of the fluid on
the sheet assuming the  no-slip boundary condition
\begin{equation}\label{BCy1}
\nabla{\psi}|_{(x_{m},y_{m})}=\epsilon b\cos{(x-t)}\textbf{e}_{x}+\epsilon a\sin{(x-t-\phi)}\textbf{e}_{y}
\end{equation}

In the far field, $y\rightarrow\infty$,  the flow moves at the unknown steady speed, $U\bm{e_x}$, whose  sign and value  will
depend on the particular movement of the material points on the sheet and on the fluid rheology. The main goal of this paper
is to derive analytically the non-Newtonian contributions to $U$.

\subsection{Constitutive equation: the Carreau model}

\begin{figure}[t]
\centering
\includegraphics[scale=0.55]{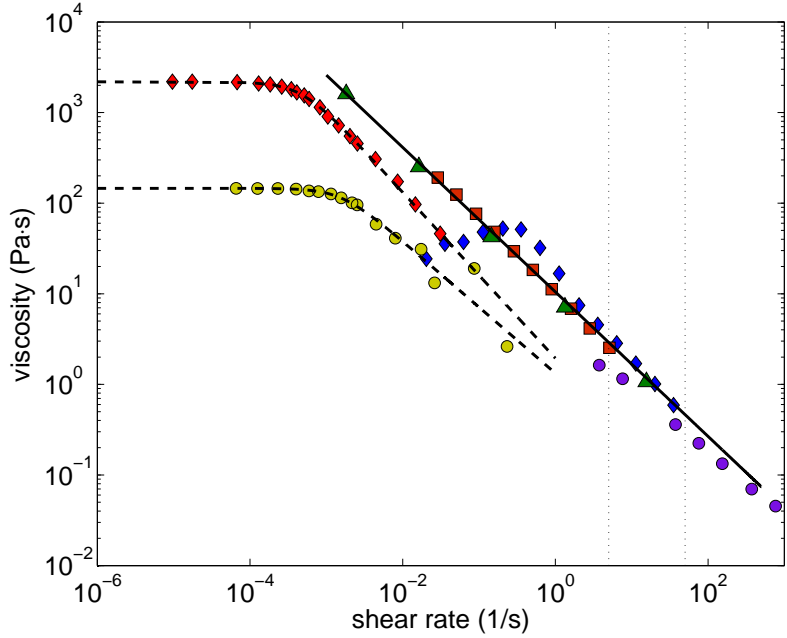}
\caption{Shear-thinning viscosity of mucus  reproduced from published data: Steady (\textcolor{blue}{$\blacklozenge$}) and
oscillatory (\textcolor{Mahogany}{$\blacksquare$}) measurements of human sputum \cite{Dawson2003}; oscillatory measurements
(\textcolor{OliveGreen}{$\blacktriangle$}) of pig small intestine mucus \cite{Sellers1991}; steady measurements
(\textcolor{RoyalPurple}{$\bullet$})  of human cervicovaginal mucus \cite{Lai2007}; micro-rheological measurements
(\textcolor{red}{$\blacklozenge$})  of human lung mucus \cite{Hwang1969}, micro-rheological measurements
(\textcolor{YellowGreen}{$\bullet$})  of human cervical  mucus \cite{Hwang1969}. When the data was reported in terms of
oscillatory moduli, the Cox-Merz rule was applied to estimate the steady values \cite{Bird1987}. The continuous line denotes
a power-law fitting, $\eta=10.5\dot{\gamma}^{-0.8}$. The dashed lines denote a fitting to the Carreau model for the lung
mucus (\textcolor{red}{$\blacklozenge$}: $\eta_{o}=2187.5, \lambda_{t}=2154.6, n=0.08$) and the cervical human mucus
(\textcolor{YellowGreen}{$\bullet$}: $\eta_{o}=145.7, \lambda_{t}=631.04, n=0.27$). The shear-rate range enclosed by the
vertical lines, $5\lesssim \dot{\gamma}\lesssim 50$ s$^{-1}$, indicates the typical oscillation frequencies of cilia
\cite{Brennen1977}. } \label{mucus}
\end{figure}

As discussed above, many biological fluids exhibit non-Newtonian rheology. In Fig.~\ref{mucus} we reproduce the data from
several published experiments on the shear-dependance of  mucus viscosity \cite{Dawson2003,Sellers1991,Lai2007,Hwang1969}.
The plot shows the value of the shear viscosity, $\eta$, as a function of the shear rate, $\dot{\bm{\gamma}}$. Measurements
are either steady-state or, for oscillatory data, their steady-state values are estimated using the Cox-Merz rule
\cite{Bird1987}.  Clearly,  mucus is very strong shear-thinning, the viscosity dropping by three orders of magnitude on the
range of shear rates $10^{-3}\lesssim\dot{\gamma}\lesssim 10^2$.

The dependance of the viscosity with the shear-rate can be modeled assuming  a Carreau generalized Newtonian fluid whose
viscosity is given by  \cite{Bird1987}
\begin{equation}\label{carreau}
\frac{\eta-\eta_{\infty}}{\eta_{o}-\eta_{\infty}}={\left[1+{(\lambda_{t}|\dot{\gamma}|)}^{2}\right]}^{\frac{n-1}{2}},
\end{equation}
where $\eta_{\infty}$ is the infinite-shear-rate viscosity, $\eta_{o}$ the zero-shear-rate viscosity, $n$ the dimensionless
power-law index, and $\lambda_{t}$ a time constant associated with the inverse of the shear-rate at which the viscosity
acquires the zero-shear-rate value. In eq.~\ref{carreau}, the  magnitude of the shear-rate tensor, $|\dot{\gamma}|$, is equal
to  $({\Pi/2})^{1/2}$, where the second invariant of the shear rate tensor, $\Pi$, is given by
\begin{equation}\label{second.invariant}
\Pi=\sum_{i=1}^{3}\sum_{j=1}^{3}\dot{\gamma}_{ij}\,\dot{\gamma}_{ji}=4{\left( \frac{\partial{u}}{\partial{x}}
\right)}^{2}+2{\left( \frac{\partial{u}}{\partial{y}}
\right)}^{2}+4\frac{\partial{u}}{\partial{y}}\frac{\partial{v}}{\partial{x}}+2{\left( \frac{\partial{v}}{\partial{x}}
\right)}^{2}+4{\left( \frac{\partial{v}}{\partial{y}} \right)}^{2}\cdot
\end{equation}

Based on the data reproduced in Fig.~\ref{mucus} it is clear that  $\eta_{o}\gg\eta_{\infty}$, and this is the limit we will
consider here. If we nondimensionalize  stresses and shear-rates by $\eta_{o}\omega$ and $\omega$, respectively, we obtain
the nondimensional form of the deviatoric stress,
\begin{equation}\label{const.relation}
\bm{\tau}=\left[{1+{(\Cu |\dot{\gamma}|)}^{2}}\right]^{\mathcal{N}}\dot{\bm{\gamma}} ,\end{equation} where we have defined
the non-Newtonian index $\mathcal{N} \equiv ({n-1})/{2}$ and where $\Cu=\omega\lambda_{t}$ is the Carreau number. When $n=1$
(equivalently, ${\cal N}=0$), the Newtonian limit is recovered in eq.~\ref{const.relation}. Clearly, other classical models
used to describe inelastic fluids may be used to fit the rheological data of mucus.  We chose the Carreau model due to its
wide range of applicability; in \ref{appendix} we discuss the class of generalized Newtonian fluids for which our analysis
remains valid. We demonstrate in particular that the model we chose allows us to formally derive the leading-order effects of
shear-dependence on the flow. Any other model, provided that it is well-defined in the zero-shear-rate limit, would either
give results mathematically equivalent to ours, or would predict an impact on the flow generated by the sheet arising at a
higher-order in the sheet amplitude.

\section{Asymptotic solution for small wave amplitudes}
\label{asymptotic} Having posed the mathematical problem, we proceed in this section to compute its solution order by order
in the dimensionless wave amplitude, $\epsilon$ \cite{Taylor1951,Childress1981,Lauga2007}.  Our   goal is to calculate the
leading-order influence of the shear-dependence of the viscosity on the flow field and, in particular, the flow induced at
infinity, $U\bm{e_x}$.  We will also compute its influence on the rate of work done by the sheet in order to deform and the
resulting transport efficiency.

We therefore write a regular perturbation expansion of the form
\begin{eqnarray}
\{\textbf{u},\psi,\bm{\tau},p,\dot{\bm{\gamma}},|\dot{\gamma}|\}&=&\phantom{+}
\epsilon \{\textbf{u}^{(1)},\psi^{(1)},\bm{\tau}^{(1)},p^{(1)},\dot{\bm{\gamma}}^{(1)},|\dot{\gamma}|^{(1)}\} \notag \\
&& +\epsilon^2\{\textbf{u}^{(2)},\psi^{(2)},\bm{\tau}^{(2)},p^{(2)},\dot{\bm{\gamma}}^{(2)},|\dot{\gamma}|^{(2)}\}\notag\\ &&
+...\label{pert.expan}
\end{eqnarray}
Using the definition of the magnitude of the shear-rate tensor, $|\dot{\gamma}|=\sqrt{\Pi/2}$, the term $|\dot{\gamma}|^{2}$
in eq.~\ref{const.relation} can be expanded in terms of $\epsilon$ as
\begin{equation}\label{expansion.magnitude}
|\dot{\gamma}|^{2}=\epsilon^{2}\frac{\Pi^{(1)}}{2}+2\epsilon^{3}|\dot{\gamma}^{(1)}||\dot{\gamma}^{(2)}|+\mathcal{O}(\epsilon^{4}).
\end{equation}
Writing the viscosity in eq.~\ref{const.relation} as
\begin{equation}\label{const.relation2}
\eta =
\left[1+\epsilon^{2}\frac{\Cu^{2}}{2}\Pi^{(1)}+2\epsilon^{3}\Cu^{2}|\dot{\gamma}^{(1)}||\dot{\gamma}^{(2)}|+\mathcal{O}(\epsilon^{4})\right]^{\mathcal{N}},
\end{equation}
and Taylor-expanding eq.~\ref{const.relation2} up to  $\mathcal{O}(\epsilon^{2})$, we finally obtain the constitutive
relationship as
\begin{equation}\label{const.relation3}
\epsilon\tau^{(1)}+\epsilon^{2}\tau^{(2)}+...=\left[1+\epsilon^{2}\frac{\mathcal{N}\Cu^{2}}{2}\Pi^{(1)}\right]\left[\epsilon\dot{\gamma}^{(1)}+\epsilon^{2}\dot{\gamma}^{(2)}+...\right].
\end{equation}

Note that, because of the $\epsilon \to - \epsilon$ symmetry, the fluid transport at infinity, quantified by $U\bm{e_x}$, has
to scale as an even power of $\epsilon$. Since the perturbation in the viscosity, eq.~\ref{const.relation2}, is order
$\epsilon^2$, this means that the flow up to order $\epsilon^2$ is going to be the same as the Newtonian one, and thus we
expect the non-Newtonian effects to come in only at order $\epsilon^4$ in $U$ (or higher, see \ref{appendix}).
 As we see below, this means that we have to
compute the flow field everywhere up to order $\epsilon^3$.

\subsection{Solution of the velocity field at $\mathcal{O}(\epsilon)$ and $\mathcal{O}(\epsilon^2)$}

Is is straightforward  to get from eq.~\ref{const.relation3} that $\bm{\tau}^{(1,2)}=\dot{\bm{\gamma}}^{(1,2)}$ and thus the
solutions at order one and two are the same as the Newtonian ones. The general procedure to derive these solutions consists
in  taking the divergence and then the curl of eq.~\ref{lowReequation} to eliminate the pressure, substituting the
constitutive relation for the stress, and using the definition of the streamfunction, eq.~\ref{incompressibility}, together
with the definition of the shear-rate, $\dot{\bm{\gamma}}=\nabla\bm{u}+\nabla\bm{u}^{T}$, to obtain the biharmonic equation
satisfied by the streamfunction
\begin{equation}\label{bih}
\nabla^{4}\psi^{(p)}=0,\quad p=1,2.
\end{equation}
The general solution of eq.~\ref{bih}  at order $\mathcal{O}(\epsilon^p)$ is already known to be
\cite{Childress1981,Leal2007}
\begin{equation}\label{biharm.homo}
\psi^{(p)}=U^{(p)}y+\sum_{j=1}^{\infty}\left[\left(A_{j}^{(p)}+B_{j}^{(p)}y\right)sin(jz)+\left(C_{j}^{(p)}+D_{j}^{(p)}y\right)cos(jz)\right]e^{-jy},
\end{equation}
which already satisfies the boundary conditions in the far field, $u=\partial\psi/\partial{y}|_{(x,\infty)}=U^{(p)}$ and
$v=-\partial\psi/\partial{x}|_{(x,\infty)}=0$.  The values of the coefficients in eq.~\ref{biharm.homo} are computed  using
the boundary conditions on the sheet, eq.~\ref{BCy1}, where $\nabla\psi$ is expanded in terms of $\epsilon$ using  a Taylor
approximation of the gradients around the material points, $x_{m}=x$ and $y_{m}=0$. The streamfunctions at
$\mathcal{O}(\epsilon)$ and $\mathcal{O}(\epsilon^2)$ are then \cite{Childress1981}
\begin{eqnarray}\label{S1}
\psi^{(1)}&=&\left(b+\beta{y}+b{y}\right)e^{-y}\sin{z}-\gamma{y}e^{-y}\cos{z},\\
\notag
\psi^{(2)}&=&\frac{1}{2}\left(b^2+2b\beta-a^2\right){y}+\frac{y}{2}e^{-2y}\left(\gamma^2-\beta^2-2\beta{b}-b^2\right)\cos 2z
\\&& -{y}e^{-2y}\left(b\gamma+\beta\gamma\right)\sin 2z  \label{S2} ,\end{eqnarray} where the net contribution to the fluid
velocity in the far field appears up to $\mathcal{O}(\epsilon^2)$ as
\begin{equation}\label{final_order2}
U^{(2)} = \frac{1}{2}\left(b^2+2a{b}\cos\phi-a^2\right).
\end{equation}

\subsection{Solution of the velocity field at $\mathcal{O}(\epsilon^3)$}
\label{solutionthirdorder}

At  order three, eq.~\ref{const.relation3} leads to   the non-Newtonian contribution from the constitutive equation as
\begin{equation}\label{S3.1}
\tau^{(3)}=\frac{\mathcal{N}\Cu^{2}}{2}\Pi^{(1)}\dot{\gamma}^{(1)}+\dot{\gamma}^{(3)}.
\end{equation}
The values of the tensor invariant, $\Pi^{(1)}$, and the components of the shear rate tensor, $\dot{\bm{\gamma}}^{(1)}$, can
be computed using the first order solution, eq.~\ref{S1}, recalling that  $\dot{\bm{\gamma}}=\nabla\bm{u}+\nabla\bm{u}^{T}$,
and using eq.~\ref{second.invariant} we obtain
\begin{eqnarray}
 u^{(1)}&=&e^{-y}\left\{ \left[\beta-(\beta+b)y \right]\sin{z}+\gamma(y-1)\cos{z} \right\} , \\
 v^{(1)}&=&-e^{-y}\left\{ \left[b+(\beta+b)y \right]\cos{z}+\gamma{y}\sin{z} \right\},
\end{eqnarray}
and therefore
\begin{eqnarray}
 \dot{\gamma}^{(3)}_{11}&=&2e^{-y}\left\{ [\beta-(\beta+b)y]\cos{z}-\gamma(y-1)\sin{z}\right \} , \\
 \dot{\gamma}^{(3)}_{12}&=&\dot{\gamma}^{(3)}_{21}=2e^{-y}\left\{ [(\beta+b)y-\beta]\sin{z}+\gamma(1-y)\cos{z}\right \} , \\
 \dot{\gamma}^{(3)}_{22}&=&-\dot{\gamma}^{(3)}_{11},
\end{eqnarray}
which allows us to finally compute
\begin{equation}\label{INV.1}
\Pi^{(1)}=8e^{-2y}\left[b^2y^2+2a{b}(y^2-y) \cos{\phi}+a^2(y-1)^2\right].
\end{equation}

After taking the divergence and then the curl of eq.~\ref{S3.1} we obtain  inhomogeneous biharmonic equation for the
streamfunction at order three given by
\begin{eqnarray}\nonumber
\nabla^{4}\psi^{(3)}&=&16\mathcal{N}\Cu^{2}e^{-3y}\left\{\left[\gamma{b^2}f(y)+2b\beta\gamma{g(y)}+a^2\gamma{h(y)}\right]\cos{z}\right.\\
&& \left.+\left[b^3i(y)-3b^2\beta{f(y)}-b(3\beta^2+\gamma^2)g(y)-a^2\beta{h(y)}\right]\sin{z}\right\} ,\end{eqnarray}
 where the $y$-dependent functions are
\begin{eqnarray}
 f(y)&=&2y^3-8y^2+7y-1 , \\
 g(y)&=&2y^3-10y^2+13y-4 , \\
 h(y)&=&2y^3-12y^2+21y-11, \\
 i(y)&=&-2y^3+6y^2-3y.
\end{eqnarray}
The homogeneous solution of this equation, $\psi_{h}^{(3)}$,  is given by the general solution in eq.~\ref{biharm.homo}. The
particular solution, $\psi_{p}^{(3)}$, can be found using the method of variation of parameters, leading to
\begin{eqnarray}\nonumber
\psi_{p}^{(3)}&=&\mathcal{N}\Cu^{2}e^{-3y}\left[\{\gamma{b^2}F(y)+2b\beta\gamma{G(y)}+a^2\gamma{H(y)}\}\cos{z}\right.\\
&& \left.+\{b^3I(y)-3b^2\beta{F(y)}-b(3\beta^2+\gamma^2)G(y)-a^2\beta{H(y)}\}\sin{z}\right] ,\end{eqnarray}
 where the $y$-dependent functions are
\begin{eqnarray}
  F(y)&=&\frac{1}{2}y^3+\frac{1}{4}y^2+\frac{1}{16}y,  \\
  G(y)&=&\frac{1}{2}y^3-\frac{1}{4}y^2+\frac{1}{16}y+\frac{1}{16},  \\
  H(y)&=&\frac{1}{2}y^3-\frac{3}{4}y^2+\frac{9}{16}y-\frac{1}{8},  \\
 I(y)&=&-\frac{1}{2}y^3-\frac{3}{4}y^2-\frac{9}{16}y-\frac{3}{16}\cdot
\end{eqnarray}

In order to determine the unknown constants in the homogeneous solution, we need to evaluate the boundary condition on the
sheet at  $\mathcal{O}(\epsilon^3)$. Using a Taylor expansion, we have
\begin{eqnarray}
\notag \nabla\psi^{(3)}|_{(x,0)}&=&-a\cos{\xi}\frac{\partial}{\partial{x}}\nabla\psi^{(2)}|_{(x,0)}-b\sin{z}\frac{\partial}{\partial{y}}\nabla\psi^{(2)}|_{(x,0)}  \\
\notag     &&-\frac{1}{2}a^2\cos^2\xi\frac{\partial^2}{\partial{x^2}}\nabla\psi^{(1)}|_{(x,0)}-a{b}\cos\xi\sin{z}\frac{\partial}{\partial{x}}\frac{\partial}{\partial{y}}\nabla\psi^{(1)}|_{(x,0)}  \\
     &&-\frac{1}{2}b^2\sin^2z\frac{\partial^2}{\partial{y^2}}\nabla\psi^{(1)}|_{(x,0)} \label{BC.3}.
\end{eqnarray}
After substituting in eq.~\ref{BC.3} the solutions at order one and two  we obtain, after some tedious but straightforward
algebra,
\begin{eqnarray}\label{BC.x}
\notag  \frac{\partial}{\partial{x}}\psi^{(3)}|_{(x,0)}&=&\frac{1}{8}\left(-3b^3-6b^2\beta+b\beta^2+3b\gamma^2\right)\cos{z}+\frac{3}{8}\left(b^3+2b^2\beta+b\beta^2-b\gamma^2\right)\cos 3z  \\
     &&-\frac{1}{4}\left(b^2\gamma+b\beta\gamma\right)\sin{z}
     +\frac{3}{4}\left(b^2\gamma+b\beta\gamma\right)\sin 3 z,
\end{eqnarray}
and
\begin{eqnarray}\label{BC.y}
 \frac{\partial}{\partial{y}}\psi^{(3)}|_{(x,0)}&=&\phantom{+}\frac{1}{8}\left(-15b^2\gamma-8b\beta\gamma\notag +3a^2\gamma\right)\cos{z}\\ && +\frac{1}{8}\left(15b^2\gamma+24b\beta\gamma\notag +9\beta^2\gamma-3\gamma^3\right)\cos 3z  \\
\notag      &&+\frac{1}{8}\left(2b^3+5b^2\beta+4b\beta^2-4b\gamma^2-3\beta\gamma^2-3\beta^3\right)\sin z \\
    & &+\frac{1}{8}\left(-6b^3-15b^2\beta-12b\beta^2+12b\gamma^2+9\beta\gamma^2-3\beta^3\right)\sin 3z .
\end{eqnarray}
Applying these boundary conditions to the general solution, $\psi^{(3)}= \psi_{h}^{(3)}+\psi_{p}^{(3)}$,  we find the value
of the only nonzero coefficients in eq.~\ref{biharm.homo} which are given by
\begin{eqnarray}
\notag     A_{1}^{(3)}&=&\frac{3}{8}\left(\frac{1}{2}\mathcal{N}\Cu^{2}-1\right)b^3+\frac{1}{8}\left(\frac{3}{2}\mathcal{N}\Cu^{2}+1\right)b\beta^2+\frac{1}{8}\left(\frac{1}{2}\mathcal{N}\Cu^{2}+3\right)b\gamma^3\\
        &&-\frac{1}{8}\left(\mathcal{N}\Cu^{2}a^2+6b^2\right)\beta,\\
     A_{3}^{(3)}&=&\frac{1}{8}b^3+\frac{1}{4}b^2\beta+\frac{1}{8}b\beta^2-\frac{1}{8}b\gamma^2,\\
\notag      B_{1}^{(3)}&=&\frac{1}{8}\left(\frac{3}{2}\mathcal{N}\Cu^{2}-1\right)b^3+\frac{1}{8}\left(\frac{3}{2}\mathcal{N}\Cu^{2}-1\right)b^2\beta+\frac{1}{8}\left(5-\frac{3}{2}\mathcal{N}\Cu^{2}\right)b\beta^2\\
          &&-\frac{1}{8}\left(1+\frac{1}{2}\mathcal{N}\Cu^{2}\right)b\gamma^2-\frac{3}{8}\beta\gamma^2+\frac{13}{16}\mathcal{N}\Cu^{2}a^2\beta-\frac{3}{8}\beta^3,\\
     B_{3}^{(3)}&=&-\frac{3}{8}b^3-\frac{9}{8}b^2\beta-\frac{9}{8}b\beta^2+\frac{9}{8}b\gamma^2-\frac{3}{8}\beta^3+\frac{9}{8}\beta\gamma^2,\\
     C_{1}^{(3)}&=&\frac{1}{4}\left(1-\frac{1}{2}\mathcal{N}\Cu^{2}\right)b\beta\gamma+\frac{1}{4}\left(\frac{1}{2}\mathcal{N}\Cu^{2}a^2+b^2\right)\gamma,\\
     C_{3}^{(3)}&=&-\frac{1}{4}\left(b^2\gamma+b\beta\gamma\right),\\
\notag      D_{1}^{(3)}&=&-\frac{1}{8}\left(\frac{1}{2}\mathcal{N}\Cu^{2}+15\right)b^2\gamma+\frac{1}{4}\left(\frac{1}{2}\mathcal{N}\Cu^{2}-3\right)b\beta\gamma\\ &&+\frac{1}{4}\left(b^2+\frac{3}{2}a^2-\frac{13}{4}\mathcal{N}\Cu^{2}a^2\right)\gamma,\\
     D_{3}^{(3)}&=&\frac{9}{8}\beta^2\gamma-\frac{3}{8}\gamma^3+\frac{9}{8}b^2\gamma+\frac{9}{4}b\beta\gamma,
\end{eqnarray}
and, as expected, the fluid velocity at infinity at this order is
\begin{equation}
     U^{(3)}=0.
\end{equation}


\begin{figure}[t]
\begin{center}
{\mbox{\subfigure[$\mathcal{N}Cu^{2}=0$,
$\phi=0$]{\includegraphics[width=0.32\columnwidth]{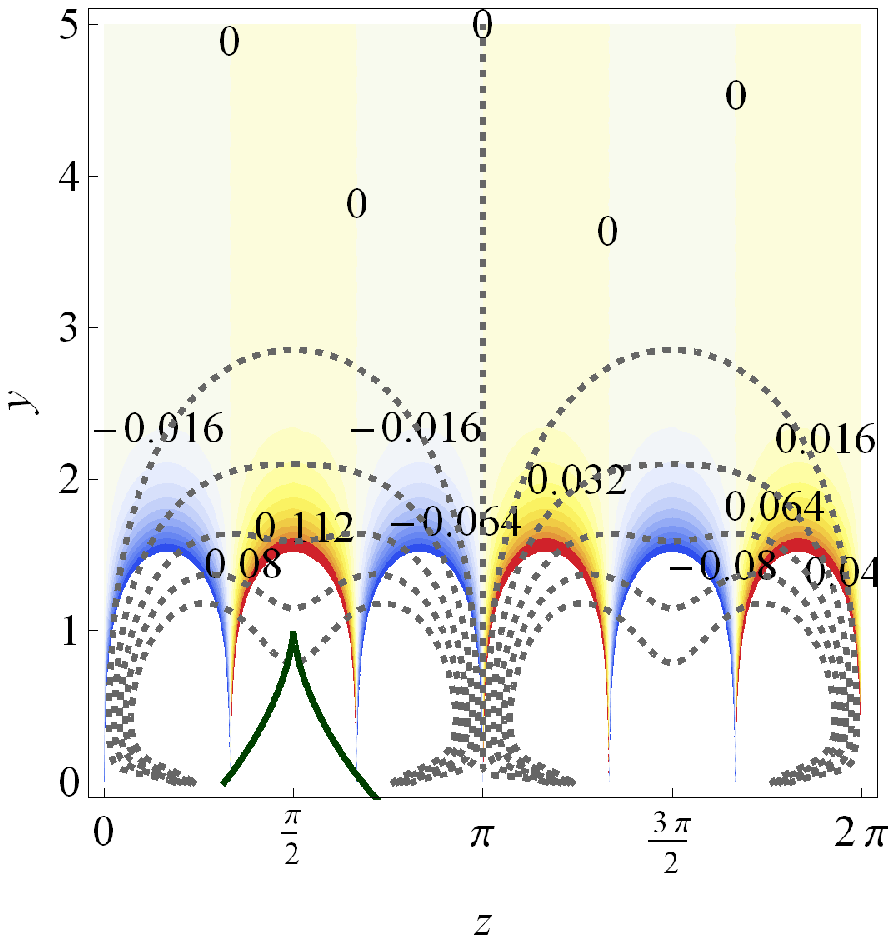}}}} {\mbox{\subfigure[$\mathcal{N}Cu^{2}=-10$, $\phi=0$]{\includegraphics[width=0.32\columnwidth]{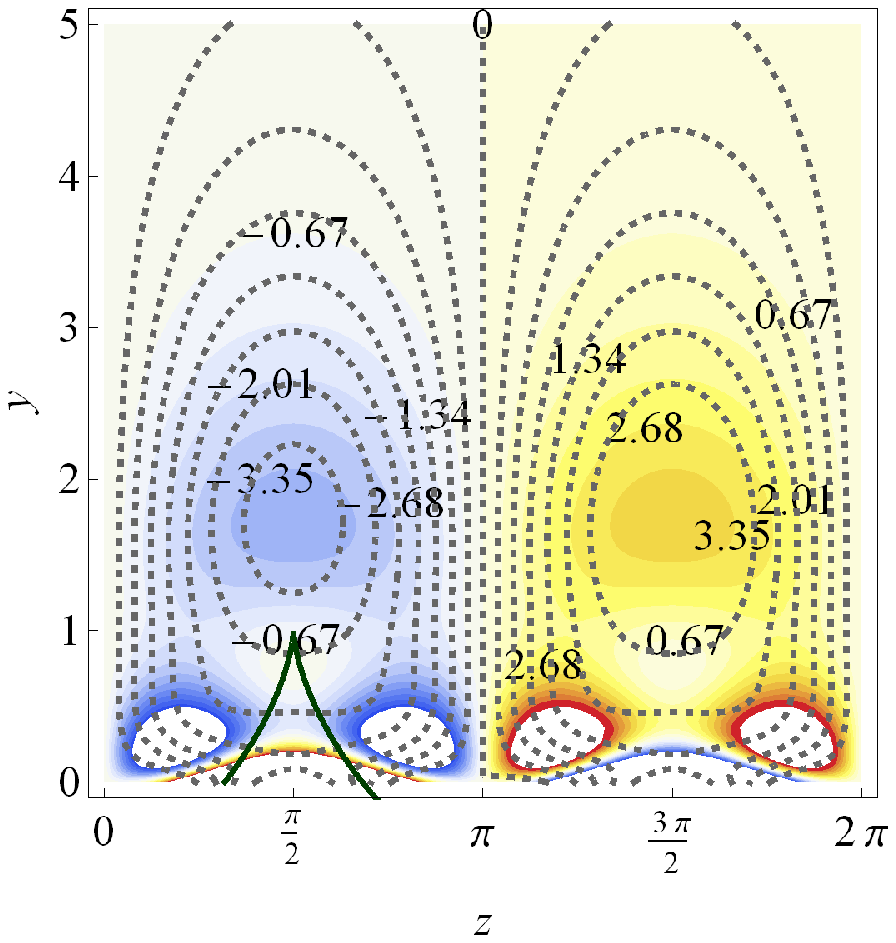}}}}
{\mbox{\subfigure[$\mathcal{N}Cu^{2}=10$, $\phi=0$]{\includegraphics[width=0.32\columnwidth]{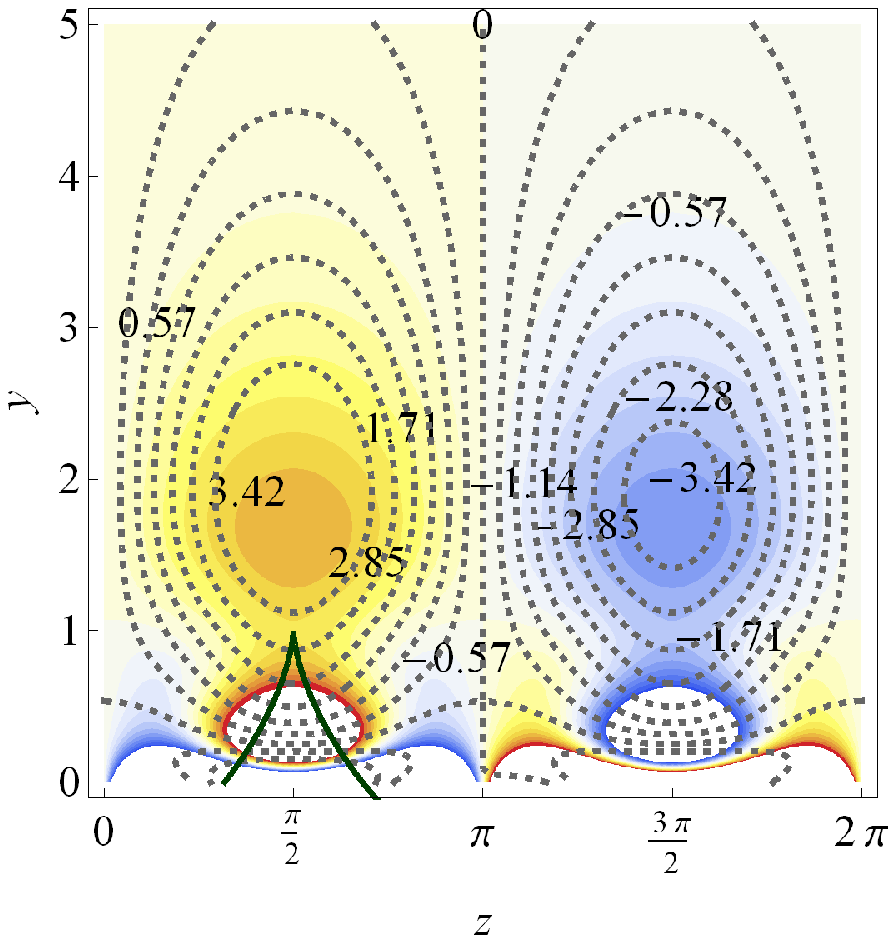}}}}\\
{\mbox{\subfigure[$\mathcal{N}Cu^{2}=0$, $\phi=\pi/4$]{\includegraphics[width=0.32\columnwidth]{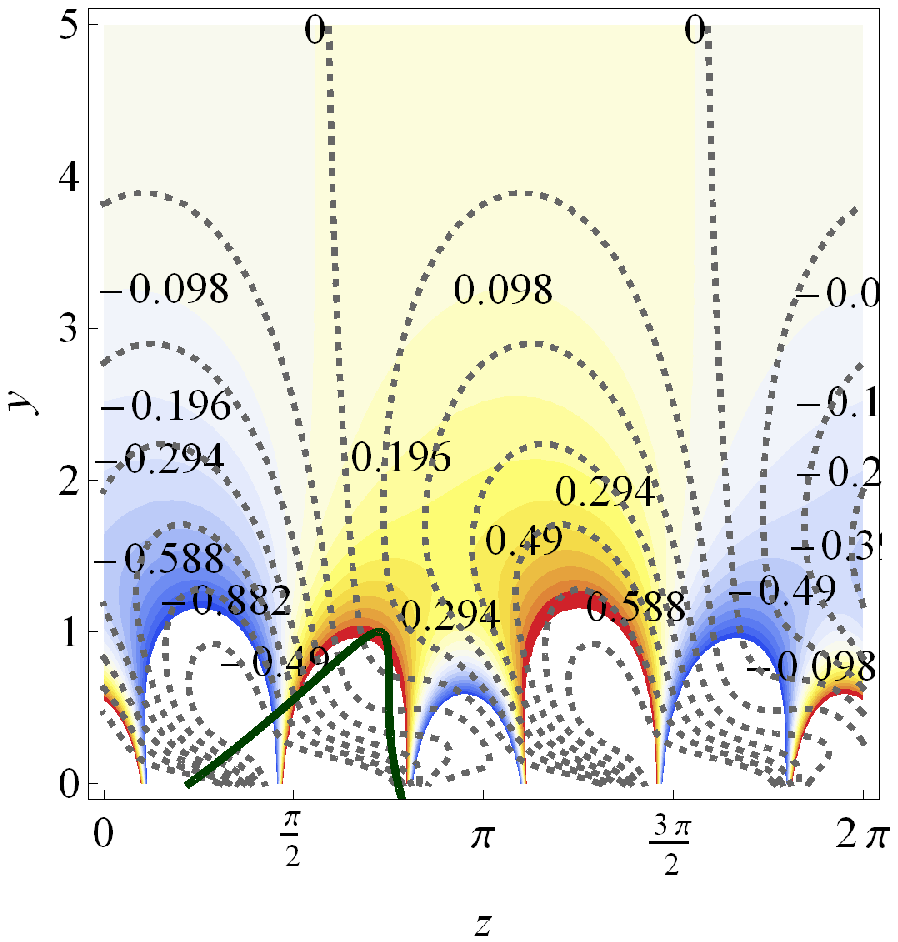}}}}
{\mbox{\subfigure[$\mathcal{N}Cu^{2}=-10$,
$\phi=\pi/4$]{\includegraphics[width=0.32\columnwidth]{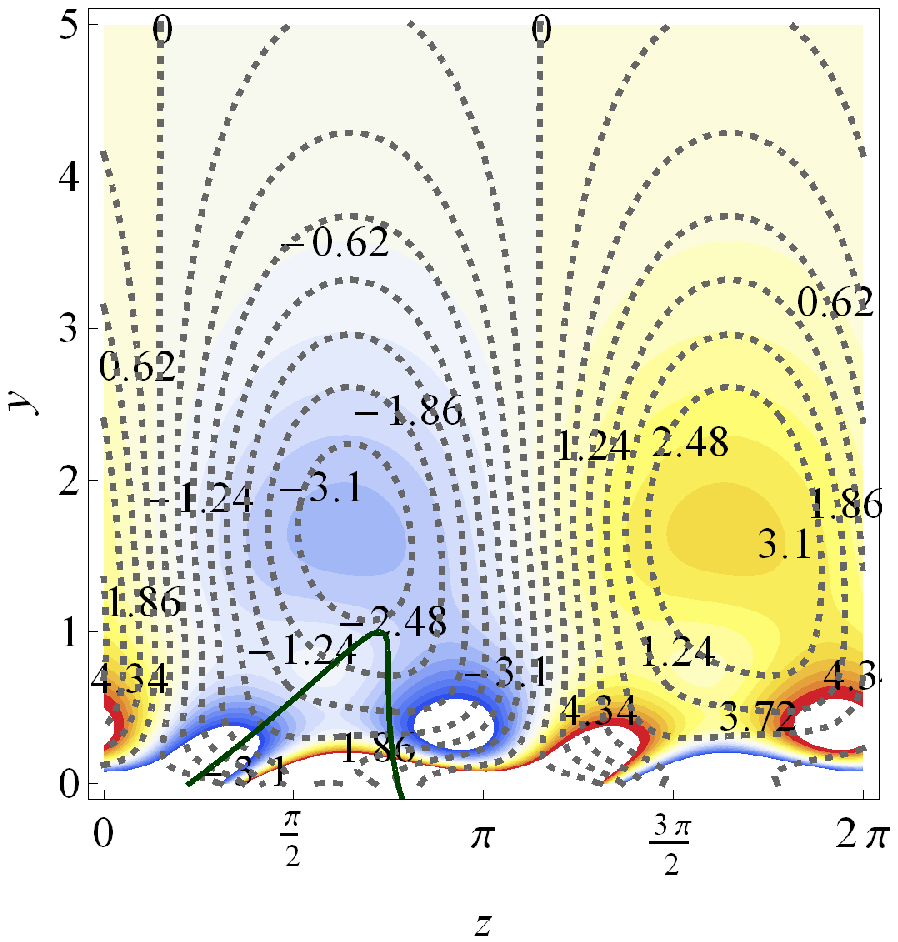}}}}
{\mbox{\subfigure[$\mathcal{N}Cu^{2}=10$,
$\phi=\pi/4$]{\includegraphics[width=0.32\columnwidth]{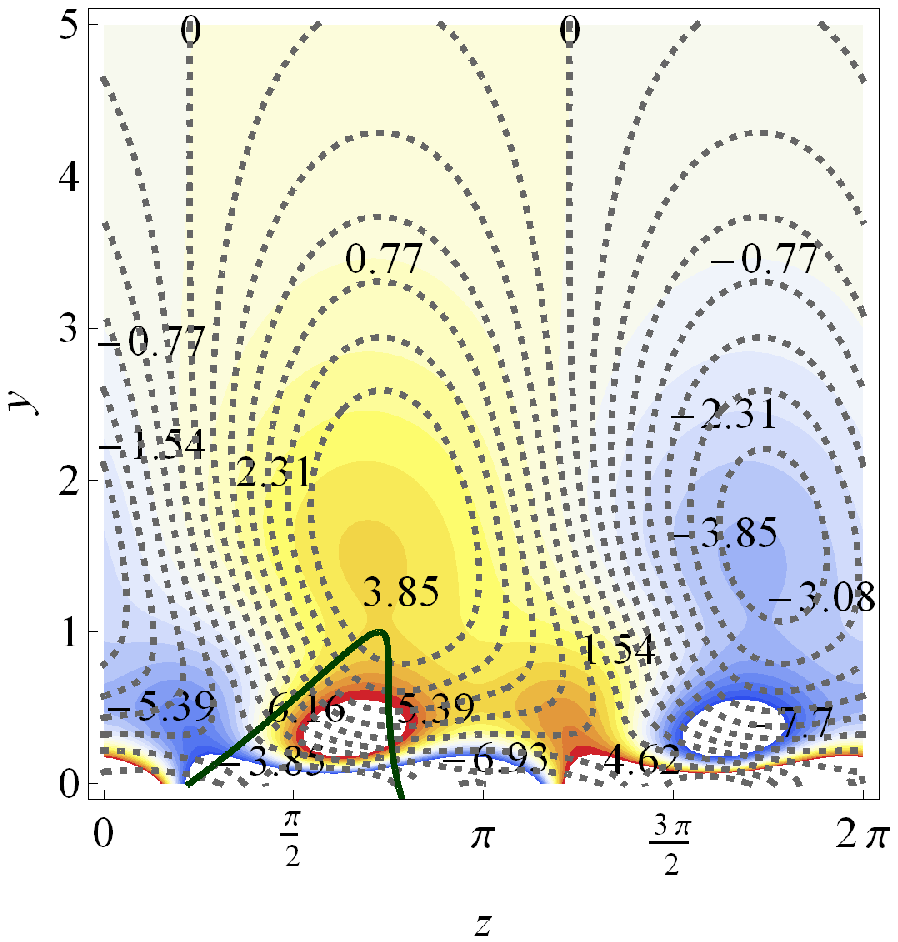}}}}\\
\end{center}
\caption{ Instantaneous streamlines (dotted lines)  and iso-vorticity contours (color levels and iso-value numbers) at order
$\epsilon^{3}$  obtained for the case $a=b$ for two values of the phase,  $\phi=0$  (top row) and $\pi/4$ (bottom row), and
three values of the  non-Newtonian term, $\mathcal{N}Cu^{2}=0$ (Newtonian fluid, left column), $-10$ (shear-thinning fluid,
middle column) and $+10$ (shear-thickening, right column).  The solid lines denote the position of the traveling wave at
$t=0$.} \label{analisis1}
\end{figure}

In order to illustrate quantitatively the difference between the Newtonian flow induced by the traveling wave and the new
non-Newtonian terms, we show in  Fig.~\ref{analisis1} the  instantaneous streamlines (dotted lines) and contours (colors and
numbers) of iso-values of the vorticity at order $\mathcal{O}(\epsilon^{3})$ for two different values of the phase, $\phi$,
and three different values of the non-Newtonian term, $\mathcal{N}Cu^{2}$, assuming $a=b$. Changing the phase $\phi$ has the
effect of tilting the vortical streamlines formed in the flow. Increasing the magnitude  of $\mathcal{N}Cu^{2}$ leads to a
significant modification of the flow:  large strong  vortices are created  above the waving sheet whose sign changes with the
sign of $\mathcal{N}Cu^{2}$. The  location and strength of the regions of high vorticity near the sheet are also modified.
Note  that even though the non-Newtonian term is scaling linearly with   $\cal N$, the total flow, sum of the Newtonian plus
the non-Newtonian component, is not anti-symmetric shear-thinning (${\cal N}<0$) vs.~shear-thickening (${\cal N}>0$), that
is, the magnitude of the values changes as well as the sign.

\subsection{Net far-field velocity at $\mathcal{O}(\epsilon^4)$}

To finish the calculation we have to compute the    velocity induced in the far field, $U^{(4)} {\bf e}_x$, to quantify the
role of non-Newtonian stresses on fluid transport, either pumping or propulsion. This can actually be determined  without
having to compute the entire flow field.  Indeed, just like for the previous order, the streamfunction,  $ \psi^{(4)}$, will
satisfy an inhomogeneous biharmonic equation. The swimming speed, which contributes to a uniform $U^{(4)} y$ term to the
streamfunction,  does not have to be computed in the far field but can instead be  evaluated using the boundary conditions.
The expansion of the $x$-velocity, $u$, at this order evaluated at $(x,0)$ is obtained by Taylor-expanding the boundary
conditions on the sheet as
\begin{eqnarray}\label{BC.4}
\notag \frac{\partial}{\partial{y}}\psi^{(4)}|_{(x,0)}&=&-b\sin{z}\frac{\partial^2}{\partial{y^2}}\psi^{(3)}|_{(x,0)}-a\cos{\xi}\frac{\partial}{\partial{x}}\frac{\partial}{\partial{y}}\psi^{(3)}|_{(x,0)}-\frac{1}{2}a^2\cos^2{\xi}\frac{\partial^2}{\partial{x^2}}\frac{\partial}{\partial{y}}\psi^{(2)}|_{(x,0)}  \\
\notag  &&-a{b}\cos{\xi}\sin{z}\frac{\partial}{\partial{x}}\frac{\partial^2}{\partial{y^2}}\psi^{(2)}|_{(x,0)}-\frac{1}{2}b^2\sin^2{z}\frac{\partial^3}{\partial{y^3}}\psi^{(2)}|_{(x,0)}\\
\notag  &&-\frac{1}{6}a^3\cos^3{\xi}\frac{\partial^3}{\partial{x^3}}\frac{\partial}{\partial{y}}\psi^{(1)}|_{(x,0)}
 -\frac{1}{2}a^2b\cos^2{\xi}\sin{z}\frac{\partial^2}{\partial{x^2}}\frac{\partial^2}{\partial{y^2}}\psi^{(1)}|_{(x,0)}\\
 &&-\frac{1}{2}a{b}^2\cos{\xi}\sin^2{z}\frac{\partial}{\partial{x}}\frac{\partial^3}{\partial{y^3}}\psi^{(1)}|_{(x,0)}-\frac{1}{6}b^3\sin^3{z}\frac{\partial^4}{\partial{y^4}}\psi^{(1)}|_{(x,0)}.
\end{eqnarray}

The first term of the right-hand side of eq.~\ref{BC.4} is the only one that provides a nonzero non-Newtonian term after
evaluating the derivatives at $y=0$. Computing the mean velocity over one period of oscillation, we obtain a Newtonian plus a
non-Newtonian contribution of the form
\begin{eqnarray}
 \label{4order.1term} \frac{1}{2\pi}\int_{0}^{2\pi} -b \sin{z}\frac{\partial^2}{\partial{y^2}}\psi^{(3)}|_{(x,0)}\,dt
 &=&
\notag \frac{1}{8}\mathcal{N}\Cu^{2}\left(3b^3\beta-6b^2\beta^2-2b^2\gamma^2-17a^2b\beta\right)\\
 &&
  +\frac{1}{16}\left(b^4+4b^3\beta+9b^2\beta^2-5b^2\gamma^2-6a^2b\beta\right).\,\,\,\,\,
\end{eqnarray}
The contribution to the mean velocity of the eight other terms appearing in the right-hand side of eq.~\ref{BC.4} are the
same as for the Newtonian solution and the final result can be found in Blake \cite{Blake1971}. Adding all the terms and
expressing the mean velocity at order four in terms of the original parameters $a$, $b$, and $\phi$, we are finally able to
write the $O(\epsilon^4)$ velocity in the far field as the sum of a Newtonian component, ${U}^{(4)}_N$, and a new
non-Newtonian term, ${U}^{(4)}_{NN}$, as
\begin{eqnarray}\label{4order.u}
{U}^{(4)}=  {U}^{(4)}_N +   {U}^{(4)}_{NN},
\end{eqnarray}
with
\begin{eqnarray}
\label{N}{U}^{(4)}_N&=&-\frac{1}{2}b^4+a^2b^2-\frac{1}{4}\left(a{b}^3+a^{3}b\right)\cos\phi,\\
    {U}^{(4)}_{NN}& = & \frac{1}{8}\mathcal{N}\Cu^{2}\left[-4a^{2}b^2+(3a{b}^3-17a^{3}b)\cos\phi-2a^{2}b^2\cos 2\phi \right].\label{NN}
\end{eqnarray}
The Newtonian component,  eq.~\ref{N}, is  Blake's solution \cite{Blake1971}. The new term, eq.~\ref{NN}, which  quantifies
the leading-order effect of a shear-dependent viscosity on the fluid transport, is the main result of our paper.

Two interesting features can be noted at this point by a simple inspection of eq.~\ref{NN}. First we see that if the sheet
does not display normal deformation ($b=0$) or no tangential deformation ($a=0$) then $    {U}^{(4)}_{NN}=0$; in order for
the shear-dependent viscosity to affect transport, it is this important that both modes of deformation be present,  $a\ne 0,
b\ne0$.  The second interesting point is that the effect scales linearly with $\mathcal{N}$, and thus with the difference
between the  power-law index, $n$, and 1. Consequently, when they have an influence on transport,  shear-thinning and
shear-thickening fluids always act in opposite direction and if one type of fluid hinders transport the other fluid will
facilitate it.

\subsection{Rate of work}
\label{work_calculation} We now turn to the calculation of the rate of work, $W$, done by the sheet to transport the fluid
(per unit area of the sheet). Since $W$ scales quadratically with the shear rate, the leading order rate of work appears at
$\mathcal{O}(\epsilon^{2})$ and is identical to the one for a Newtonian fluid \cite{Blake1971,Childress1981}. Using the
nondimensionalization $W\sim \eta_{o}\omega^{2}/k$  we have  in dimensionless variables
\begin{equation}
\langle W^{(2)} \rangle =\epsilon^{2}\left(a^2+b^2\right),
\end{equation}
where we have used  $\langle ... \rangle$ to denote mean over a period of oscillation.

In order to compute the rate of work at higher order we recall that, in the absence of inertia and without the presence of
external forces, the rate of work $W$ done by a waving surface $S$ against a fluid is equal to the value of the energy
dissipation over the enclosed volume $V$
\begin{equation}\label{work.3}
W=\int_{S}\bm{u}\cdot\bm{\sigma}\cdot\bm{n}\,dS=\int_{V}\bm{\sigma}:\nabla\bm{u}\,dV.
\end{equation}
Using that the stress tensor and velocity vector have been expanded in powers of the wave amplitude $\epsilon$, the rate of
work at third and fourth order are
\begin{eqnarray}\label{work.third}
W^{(3)}&=&\int_{V}\left(\bm{\sigma}^{(1)}:\nabla\bm{u}^{(2)}+\bm{\sigma}^{(2)}:\nabla\bm{u}^{(1)}\right)\,dV,\\
\label{work.fourth}
W^{(4)}&=&\int_{V}\left(\bm{\sigma}^{(1)}:\nabla\bm{u}^{(3)}+\bm{\sigma}^{(2)}:\nabla\bm{u}^{(2)}+\bm{\sigma}^{(3)}:\nabla\bm{u}^{(1)}\right)\,dV.
\end{eqnarray}

At first and second orders in $\epsilon$, the constitutive equations for the stress tensor are the Newtonian ones,
$\bm{\sigma}^{(1,2)}=-p^{(1,2)}\bm{I}+\dot{\bm{\gamma}}^{(1,2)}$. The constitutive equation at third order is in contrast
$\bm{\sigma}^{(3)}=-p^{(3)}\bm{I}+\frac{\mathcal{N}\Cu^{2}}{2}\Pi^{(1)}\dot{\bm{\gamma}}^{(1)}+\dot{\bm{\gamma}}^{(3)}$.
Substituting these into eqs.~\ref{work.third} and \ref{work.fourth} we see that the pressure at any order multiplies the
divergence of the velocity which is zero by incompressibility. The mean rates of work per unit area of the sheet  at order
$\mathcal{O}(\epsilon^{3})$ and $\mathcal{O}(\epsilon^{4})$ are thus given by
\begin{eqnarray}\label{work.third2}
\langle{W}^{(3)}\rangle&=&\frac{1}{2\pi}\int_{0}^{2\pi}\int_{0}^{\infty}\left(\dot{\bm{\gamma}}^{(1)}:\dot{\bm{\gamma}}^{(2)} \right) \,dydz,\\
\label{work.fourth2}
\langle{W}^{(4)}\rangle&=&\frac{1}{2\pi}\int_{0}^{2\pi}\int_{0}^{\infty}\left(\dot{\bm{\gamma}}^{(1)}:\dot{\bm{\gamma}}^{(3)}+\frac{1}{2}\dot{\bm{\gamma}}^{(2)}:\dot{\bm{\gamma}}^{(2)}+\frac{\mathcal{N}\Cu^{2}}{4}\Pi^{(1)}\dot{\bm{\gamma}}^{(1)}:\dot{\bm{\gamma}}^{(1)}\right)\,dydz.\quad\,\,\,\,
\end{eqnarray}
Substituting the values of the shear-rate tensor and its second invariant at the respective orders in $\epsilon$, we obtain
that $\langle{W}^{(3)}\rangle=0$, which could have been anticipated by symmetry $\epsilon\to -\epsilon$. The mean rate of
work per unit area at $\mathcal{O}(\epsilon^{4})$ is found to be  the sum of a Newtonian term plus a non-Newtonian
contribution as
\begin{eqnarray}\label{4work.final}
  \langle{W}^{(4)}\rangle =   \langle{W}^{(4)}\rangle_{N}+   \langle{W}^{(4)}\rangle_{NN},
\end{eqnarray}
where
\begin{eqnarray}\label{4work.final}
  \langle{W}^{(4)}\rangle_N&=&-\frac{1}{4}(a^4+b^4)+5a^{2}b^2+(a{b}^3+3a^{3}b)\cos\phi-\frac{1}{2}a^{2}b^2\cos 2\phi, \label{N_W} \\
  \langle{W}^{(4)}\rangle_{NN}&=&\frac{1}{8}\mathcal{N}\Cu^{2}\left(15a^4+3b^4+4a^{2}b^2-8a^{3}b\cos\phi+2a^{2}b^2\cos 2\phi \right).\label{NN_W}
\end{eqnarray}
Similarly to the non-Newtonian contribution to the velocity, eq.~\ref{NN}, the effect of the shear-dependent viscosity on $W$
scales linearly with   $\mathcal{N}$, and thus shear-thinning and shear-tickening fluids always contribute in opposite sign
to the rate of working of the sheet. In contrast however with the fluid velocity, we see that modes with pure normal  ($a=0$)
or   tangential ($b=0$) motion do contribute to the rate of work at this order.

\subsection{Transport efficiency}
We finally consider the efficiency of transport, ${\cal E}$, measured as the ratio between the useful work done in the fluid
to create the flow at infinity and the total dissipation, written  ${\cal E} \sim  U^2/\langle W\rangle$ in  dimensionless
variables \cite{Lauga2009}. Given that we have $U =  \epsilon^2 U^{(2)} + \epsilon^4 U^{(4)}$ and $ \langle{W}\rangle =
\epsilon^2   \langle{W}^{(2)}\rangle + \epsilon^4  \langle{W}^{(4)}\rangle$, it is clear what we will obtain ${\cal E}  =
\epsilon^2 {\cal E}^{(2)} + \epsilon^4 {\cal E}^{(4)} $ with each term found by Taylor expansion as
\begin{equation}
{\cal E}^{(2)} = \frac{\left(U^{(2)}\right)^2}{  \langle{W}^{(2)}\rangle} ,\quad
 {\cal E}^{(4)} =\frac{\left(U^{(2)}\right)^2}{  \langle{W}^{(2)}\rangle}
 \left(2\frac{U^{(4)}}{U^{(2)}}- \frac{\langle{W}^{(4)}\rangle}{\langle{W}^{(2)}\rangle}\right).
\end{equation}
The first term, ${\cal E}^{(2)}$, is the same as the Newtonian one whereas the second one, ${\cal E}^{(4)}$, includes a
non-Newtonian contribution, given by
\begin{equation}\label{eff_final}
 {\cal E}^{(4)}_{NN} =\frac{\left(U^{(2)}\right)^2}{  \langle{W}^{(2)}\rangle}
 \left(2\frac{U^{(4)}_{NN}}{U^{(2)}}- \frac{\langle{W}^{(4)}\rangle_{NN}}{\langle{W}^{(2)}\rangle}\right) \cdot
\end{equation}
Non-Newtonian stresses will lead to an increase in the efficiency of transport if the term on the right-hand side of
eq.~\ref{eff_final} is positive, which we will investigate below for the various kinematics considered.

\section{Do shear-thinning fluids facilitate locomotion and transport?}
\label{analysis}

With our mathematical results we are now ready to address the impact of non-Newtonian stresses on flow transport. We focus on
four  aspects. We first show that, within our mathematical framework, shear-thinning always decrease the cost of flow
transport. We then demonstrate that the maximum energetic gain is always obtained for antiplectic metachronal waves. When the
sheet is viewed as a model for flagellar locomotion in complex fluids, for example spermatozoa, we then show that, at the
order in which we carried out the calculations, there is   no influence of the shear-dependent viscosity on the kinematics of
swimming. We finally address the more general case of viewing the sheet as an envelope for the  deformation of cilia array,
and address the impact of non-Newtonian stresses  on the magnitude of the flow induced in the far field for a
variety of sheet kinematics.

\subsection{Shear-thinning fluids always decrease the cost of transport}
We start by considering the non-Newtonian contribution to the rate of working by the sheet, eq.~\ref{NN_W}. We can write  $
\langle{W}^{(4)}\rangle_{NN}= \mathcal{N}\Cu^{2} F(a,b,\phi)$ where the function $F$ is
\begin{equation}\label{FF}
F(a,b,\phi)=
  \frac{1}{8} \left(15a^4+3b^4+4a^{2}b^2-8a^{3}b\cos\phi+2a^{2}b^2\cos 2\phi \right).
\end{equation}
We are going to show that this function is always positive, meaning that for shear-thinning fluids (${\cal N }< 0$), the
sheet expends less energy than in a Newtonian fluid with the same zero-shear rate viscosity.

In order to show that $F\geq 0$ we  calculate its minimum value and show that it is positive. The derivative of
$F$ with respect to the phase is given by
\begin{equation}\label{d}
\frac{\d F}{\d \phi}=a^2b\sin\phi (a-b\cos\phi),
\end{equation}
while its second derivative is
\begin{equation}\label{d2}
\frac{\d^2 F}{\d \phi^2}=a^2b (a\cos\phi - b \cos 2\phi).
\end{equation}
The local extrema of $F$ are reached when ${\d F}/{\d \phi}=0$ which occurs at three points: $\phi=0, \pi$ and, when $a<b$,
$\cos\phi=a/b$. When $a \geq b$, only the points $\phi=0, \pi$ are extrema, and plugging them into eq.~\ref{d2} we see that
the minimum of $F$ is obtained at $\phi=0$ (positive second derivative). From eq.~\ref{FF} we get that this minimum value is
$F_{\rm min}=[7a^4+3b^4+6a^{2}b^2+8a^3(a-b)]/8$ which is positive since $a\geq 0$, $b\geq 0$, and $a\geq b$. In the case
where   $a < b$ it is straightforward to show that the minimum of $F$ occurs at the phase $\phi$ satisfying $\cos\phi=a/b$.
Plugging that value of the phase into eq.~\ref{FF}, and using that $\cos2\phi=2a^2/b^2-1$ we obtain that $F_{\rm min}=[11a^4
+ 3 b^4 + 2 a^2 b^2]/8$ which is also always positive.

We therefore always have $F \geq 0$ and the non-Newtonian contribution to the rate of working by the sheet, $
\langle{W}^{(4)}\rangle_{NN}$, always has the same sign as $\cal N$. For shear-thinning fluids ${\cal N} < 0$ and therefore
the cost of fluid transport is always reduced -- a result which remains true even if the non-Newtonian contribution to the
flow at infinity, $U^{(4)}_{NN}$, is equal to zero.  As a consequence, the last term in eq.~\ref{eff_final},
$\langle{W}^{(4)}\rangle_{NN}/\langle{W}^{(2)}\rangle$, will always be negative for shear-thinning fluids, indicating a
contribution to an increase in the efficiency. The opposite is true for shear-thickening fluids which always add to the
energetic cost. This result is  reminiscent of earlier work on the cost of transport by gastropod showing that shear-thinning
fluids are advantageous, although for a completely different reason (mechanical work versus mucus production)
\cite{Lauga2006}.

As a final note, it is worth pointing out that, had the flow kinematics  been unchanged at all orders compared to the
Newtonian ones, then this result of reduction in $W$ would have in fact been obvious. Indeed,  if the shear is the same
everywhere in the fluid, and if the viscosity decreases, then the total dissipation should also decrease. However, as
detailed in \S \ref{work_calculation}, the dissipation rate at order four in $\epsilon$ depends on the third-order kinematics
in the fluid which are different from what they would have been in the absence of shear-dependence. The fact that our
calculations show that shear-thinning fluids always decrease the cost of transport is therefore not  obvious.

\subsection{Antiplectic metachronal waves  offer the largest  non-Newtonian energy saving}

Another general result can be gained from a close inspection of the function $F$ in eq.~\ref{FF}. For given values of both
amplitudes $a$ and $b$, it is clear that $F$ it is maximized when $\cos \phi=-1$ and $\cos 2\phi=1$, which occurs when
$\phi=\pi$. This is thus the value of the phase between normal and tangential motion leading to the maximum non-Newtonian
decrease in the energy expanded by the sheet to transport the fluid. For this particular value of the phase difference, the
kinematics of the material point at $x=0$, is    $x_{m}= - \epsilon{a}\cos{t}$,  $y_{m}= - \epsilon{b}\sin{t}$ which means
that material points follow elliptical trajectories of semi axes $\epsilon a$ and $\epsilon b$ in the anticlockwise
direction. Viewing the sheet as a model for the motion of cilia tips this is therefore a wave for which the effective stroke
is in the $-x$ direction while the recovery stroke occurs in the $+x$ direction. Given that the metachronal waves are assumes
to propagate in the positive direction, this means that antiplectic metachronal waves  offer the largest  non-Newtonian
energy saving in a shear-thinning fluid.

\subsection{The sheet as a model for flagellar locomotion}

We have seen in  \S\ref{asymptotic} that the presence of non-Newtonian stresses affects the fluid transported by the waving
sheet only if both oscillatory motions, normal ($b\neq 0$) and tangential ($a\neq 0$), are present. In the frame moving with
the fluid at infinity, this induced speed can  be also interpreted as the swimming speed of a free-moving sheet, which
prompted Taylor's original idea to propose it as the simplest model of flagellar locomotion \cite{Taylor1951}. The difference
however between the waving kinematics of a flagellum and the one we have studied in this paper is that a flagellum is
inextensible. Consequently, the flagellum has $O(\epsilon)$ motion only in the normal direction, while the tangential
deformation is $O(\epsilon^2)$ as a  result from the requirement of inextensibility (material points on a flagellum describe
a typical figure-8 motion). Higher-order kinematics can also be described similarly \cite{Taylor1951,Childress1981,Sauzade}.

To address the impact of a shear-dependent viscosity on locomotion we have followed the same procedure as above and computed
the swimming velocity systematically up to  $\mathcal{O}(\epsilon^{4})$ using the Carreau model and enforcing inextensibility
condition. We obtain formally that $U^{(4)}_{NN}=0$: the shear-dependence of the viscosity does not affect the locomotion
speed of the sheet at this order. Inextensible swimmers, such as those employing flagella, and deforming in a constant,
small-amplitude waving motion of the type modeled here are thus expected to be hardly affected by the shear-dependence of the
fluid. Since the rate of working decreases for shear-thinning fluids, swimmers are thus more efficient than in Newtonian
fluids (see eq.~\ref{eff_final}). Our  theoretical result is consistent with  experiments by Shen and Arratia \cite{Shen2011}
which have shown that the swimming velocity of a nematode (\emph{C.~elegans}) waving freely in a shear-thinning fluid
(xanthan gum solution) is the same as that observed in a Newtonian fluid with similar viscosity.

\subsection{The sheet as a model for fluid transport and pumping}

\begin{figure}[t]
\centering
\includegraphics[scale=0.28]{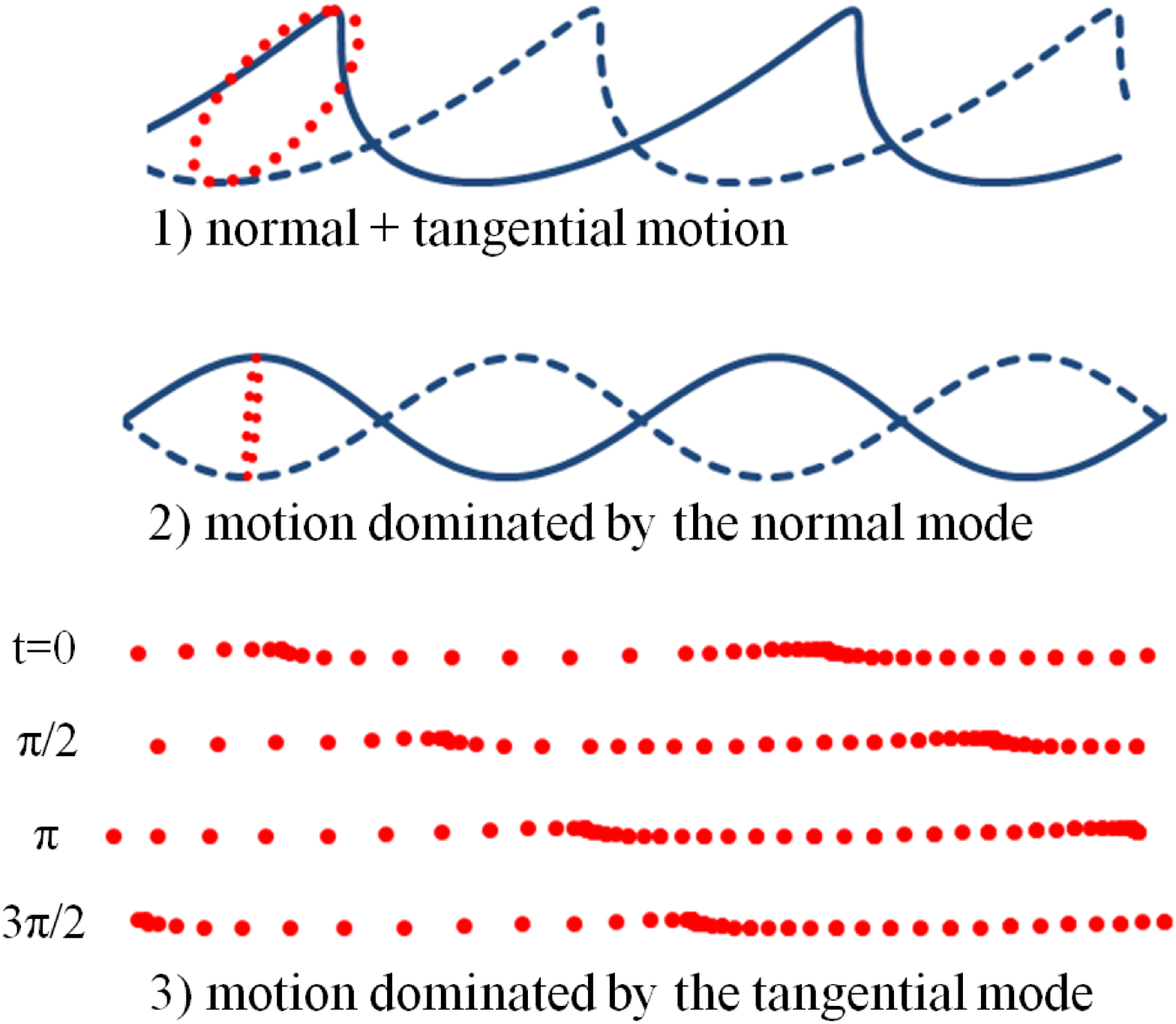}
\caption{Representation of the sheet kinematics for the three different types of motion considered in this work: (1) normal
plus tangential motion of the same magnitude, $a=b$; (2) motion dominated by the normal mode, $b\gg{a}$; (3) motion dominated
by the tangential mode, $a\gg{b}$. In all cases displayed a value of $\phi=\pi/4$ was chosen and the red dots indicate the
position of a material point during a  stroke cycle. In (1) and (2)  the position of the wave is shown at time $t=0$ (solid
line) and $\pi$ (dashed solid line).  In (3) the positions of the material points are shown at $t=0$, $\pi/2$, $\pi$, and
$3\pi/2$. } \label{wavetype}
\end{figure}

Returning to the sheet as a model for fluid transport by the tips of cilia arrays, we investigate here the effect of the
shear-dependent viscosity on the pumping performance as a function of the values  of  $a$, $b$, and $\phi$.  Recall that the
magnitude of the non-Newtonian contribution, $U^{(4)}_{NN}$,  scales with the  absolute value of $\mathcal{N}\Cu^{2}$ while
its sign is that of $\mathcal{N}$, which is positive if the fluid is shear-thickening, and negative is the fluid is
shear-thinning. Among the many combinations of $a$, $b$, and $\phi$ that we may choose, we limit the study to three relevant
cases: $a=b$, $b\gg{a}$, and $a\gg{b}$, and in each case consider four illustrative values of $\phi$, namely, $\phi=0$,
$\phi=\pm\pi/2$, and $\phi=\pi$. In Fig.~\ref{wavetype} we display  the three different kinematics of the sheets in the case
where the phase is $\phi=\pi/4$. Below we reproduce the results for the fourth-order swimming speed, $U^{(4)}$, and discuss
its relation to the leading-order Newtonian contribution at order $\epsilon^{2}$. We also compute the mean rate of working,
$\langle W^{(4)}\rangle$. The main results are summarized in Table~\ref{tabletwo} in the discussion section of the paper.

\subsubsection{Fluid transport with both normal and  tangential motion}

In the case where the normal and tangential amplitudes are similar,  $a=b\equiv c$,  we obtain
\begin{eqnarray}\label{ana.vel1}
  &\mbox{(i)} &U^{(4)} =-\frac{10c^4}{4}\mathcal{N}\Cu^{2}\mbox{, for}\, \phi=0\mbox{; the $O(\epsilon^2)$ velocity is}\,  U^{(2)}=c^2;\\
  &\mbox{(ii)}& U^{(4)}=\frac{c^4}{2}\left(1-\frac{1}{2}\mathcal{N}\Cu^{2}\right)\mbox{, for}\,
  \phi=\pm\pi/2\mbox{; }\, U^{(2)}=0; \\
  &\mbox{(iii)}& {U^{(4)}}=c^4\left(1+\mathcal{N}\Cu^{2}\right) \mbox{, for}\, \phi=\pi\mbox{; }\, U^{(2)} =-c^2.
\end{eqnarray}
We  see that for all the values of the phase $\phi$, the fluid transport  is increased, i.e.~the second order velocity and
the non-Newtonian contribution have the same sign, if the fluid is shear-thinning, ${\cal N}< 0$. This results remains true
for $\phi=\pm\pi/2$ when the mean velocity at $\mathcal{O}(\epsilon^{2})$ is zero as in this case the non-Newtonian
contribution has the same sign as the Newtonian term at order four when ${\cal N}< 0$.

On the other hand, the mean rate of work per unit length becomes
\begin{eqnarray}\label{ana.work1}
  &\mbox{(i)}& {{\langle}W^{(4)}\rangle}=c^4\left(8+2\mathcal{N}\Cu^{2}\right)\mbox{, for}\, \phi=0; \\
  &\mbox{(ii)}&  {{\langle}W^{(4)}\rangle}=c^4 \left(5+\frac{5}{2}\mathcal{N}\Cu^{2}\right)\mbox{, for}\, \phi=\pm\pi/2; \\
  &\mbox{(iii)}& {{\langle}W^{(4)}\rangle}=4c^4\mathcal{N}\Cu^{2}\mbox{, for}\, \phi=\pi;
\end{eqnarray}
and the maximum reduction of the rate of work is given when $\phi=\pi$, as anticipated above. In all cases, the transport
efficiency in eq.~\ref{eff_final} is increased for two reasons: more fluid is being transported, and transport occurs at a
lower cost.

\subsubsection{Motion dominated by the normal mode}

In the case where the sheet kinematics is dominated by the normal deformation, $b\gg{a}$,   we have $U^{(2)} \approx
{b^2}/{2}$. Keeping all terms in $a/b$ up to the leading-order non-Newtonian contribution, we obtain that the  velocity at
$\mathcal{O}(\epsilon^{4})$ is given by
\begin{eqnarray}\label{ana.vel2}
  &\mbox{(i)}& {U^{(4)}}\approx -\frac{b^4}{2}+\frac{1}{4}{ab^3}\left(\frac{3}{2}\mathcal{N}\Cu^{2}-1\right)\mbox{, for}\, \phi=0;\\
  &\mbox{(ii)}& {U^{(4)}}\approx -\frac{b^4}{2} + a^2b^2\left(1-\frac{1}{4}{\cal N}\Cu^2\right)
  \mbox{, for}\, \phi=\pm\pi/2; \\
  &\mbox{(iii)} & {U^{(4)}}\approx -\frac{b^4}{2}+\frac{1}{4} {ab^3}\left(1-\frac{3}{2}\mathcal{N}\Cu^{2}\right)\mbox{, for}\,\phi=\pi;
\end{eqnarray}
while at leading-order in $a/b$ we have
\begin{equation}\label{W4bb}
{{\langle}W^{(4)}\rangle}\approx \frac{b^4}{4}\left(\frac{3}{2}\mathcal{N}\Cu^{2}-1\right),
\end{equation}
for all values of the phase.  In the cases where $\phi=\pm\pi/2$, and $\pi$, the velocity induced by the motion of the sheet
is increased if the fluid is shear-thinning. In contrast, for in-phase motion, $\phi=0$, the non-Newtonian term acts in the
direction opposite to the leading-order Newtonian term, $U^{(2)}$. However, the non-Newtonian contribution to the velocity,
eq.~\ref{ana.vel2},  scales as $\sim {\cal N}a b^3$ while the contribution to the energetics, eq.~\ref{W4bb}, scales as
${\cal N} b^4$, and thus in the limit $b \gg a$ considered here,  eq.~\ref{eff_final} leads to a positive value for ${\cal
E}^{(4)}_{NN}$:   the motion is still more efficient in a shear-thinning fluid.


\begin{figure}[t]
\begin{center}
\includegraphics[scale=0.345]{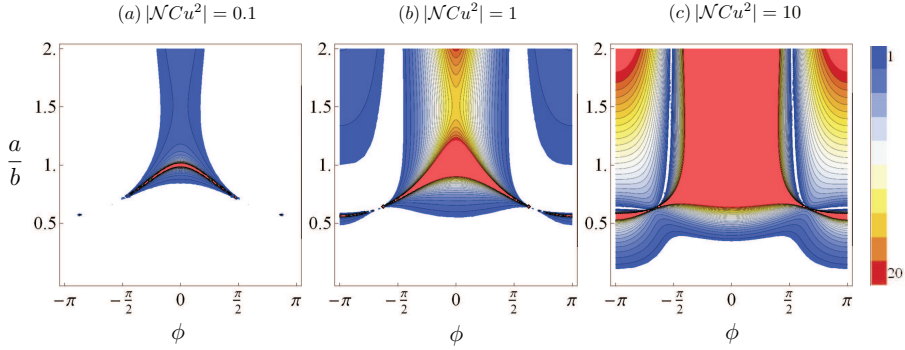}
\end{center}
\caption{ In the wave configuration space $a/b$ vs.~$\phi$ we use colors to indicate the  regions in which the magnitude of
the net non-Newtonian flow dominates that of the Newtonian component at order $\mathcal{O}(\epsilon^{4})$,
i.e.~$|U_{NN}^{(4)}|>|U_{N}^{(4)}|$ (from eqs.~\ref{N} and \ref{NN}) for three illustrative values of the non-Newtonian
coefficient, $\mathcal{N}Cu^{2}$ (0.1, 1 and 10). The cut-off value of $|U_{NN}^{(4)}|/|U_{N}^{(4)}|=20$ is chosen to account
for the zeros of $U_{N}^{(4)}$. The color scheme, shown on the right, gives the iso-values of
$|U_{NN}^{(4)}|/|U_{N}^{(4)}|$.} \label{analisis2}
\end{figure}

 \subsubsection{Motion dominated by the tangential mode}

The last case we consider is the one where $a\gg{b}$ and the kinematics is dominated by the tangential deformation. In this
case, the second order velocity is $U^{(2)} \approx -{a^2}/{2}$. Keeping all terms in $b/a$ up to the leading-order
non-Newtonian contribution in $U^{(4)}$ we obtain
\begin{eqnarray}\label{ana.vel3}
\label{U4aa}  &\mbox{(i)}& {U^{(4)}}\approx -\frac{1}{4}{ba^3}\left(1+\frac{17}{2}\mathcal{N}\Cu^{2}\right)\mbox{, for}\, \phi=0;\\
\label{U4aaa}  &\mbox{(ii)}& {U^{(4)}}\approx a^2b^2 \left(1-\frac{1}{4}{\cal N} \Cu^2\right)\mbox{, for}\, \phi=\pm\pi/2; \\
  &\mbox{(iii)}& {U^{(4)}}\approx \frac{1}{4}{b}{a^3}\left(1+\frac{17}{2}\mathcal{N}\Cu^{2}\right)\mbox{, for}\,\phi=\pi;
\end{eqnarray}
while the rate of work is given by
 \begin{equation}\label{W4aa}
{{\langle}W^{(4)}\rangle}\approx \frac{a^{4}}{4}\left(\frac{15}{2}\mathcal{N}\Cu^{2}-1\right)\cdot
\end{equation}
This time, for a shear-thinning fluid, the non-Newtonian contribution acts in the direction of the leading-order Newtonian
flow only for $\phi=\pi$ while a flow  in the opposite direction is created for $\phi=0$ and $\pm\pi/2$. However, similarly
to the discussion in the previous section, since the non-Newtonian effect on the rate of  working is $\sim {\cal N}a^4$
(eq.~\ref{W4aa}) while the impact on the flow speed is   $\sim {\cal N}ba^3$  (eq.~\ref{U4aa}) and $\sim {\cal N}a^2b^2$
(eq.~\ref{U4aaa}), the transport efficiency, eq.~\ref{eff_final}, is   systematically decreased in the shear-thinning case.

\subsubsection{Numerical approach}
\label{numerics}

With the solution for both the net Newtonian flow speed, $U_{N}^{(4)}$  (eq.~\ref{N}),  and the non-Newtonian one,
$U_{NN}^{(4)}$ (eqs.~\ref{NN}), it is straightforward to   numerically compare their  values. In Fig. \ref{analisis2} we plot
the regions in which the ratio in magnitude between both terms, $|U_{NN}^{(4)}/U_{N}^{(4)}|$, is above 1 in the  wave
configuration space $a/b$ vs.~$\phi$, for three values of the non-Newtonian term $\mathcal{N}Cu^{2}$ (0.1, 1, and 10). The
figure indicates therefore the regions where the non-Newtonian contribution dominates the Newtonian one.
  The color scheme for the iso-values of the velocity ratio is indicated on the right of the figure, and the  cut-off of the ratio  to account for the zeros of $U_{N}^{(4)}$ is chosen to be 20.  When the flow is almost Newtonian ($\mathcal{N}Cu^{2}=0.1$) the non-Newtonian contribution can almost everywhere be neglected, except near the region where the Newtonian term exactly cancels out. As $\mathcal{N}Cu^{2}$ increases, the non-Newtonian contribution to the net flow becomes predominant over almost all  configuration space; the only region remaining white (indicating a ratio of less than 1) is at the bottom of each  panel indicating the prevalence of the Newtonian term for waves dominates by normal modes ($a/b \ll 1$).

\section{Discussion}
\label{discussion}

In this paper we propose a mathematical modeling approach  to  address the role of  shear-dependent  viscosity on flagellar
locomotion and transport of mucus by cilia array. We employ the  envelope model originated by  Taylor
\cite{Taylor1951,Childress1981} and allow for both normal and tangential deformation of a two-dimensional waving sheet. We
compute the flow field induced by a small-amplitude deformation of the envelope in a generalized Newtonian Carreau fluid with
power index $n$ up to order 4 in the dimensionless waving amplitude, $\epsilon$. The net flow induced at infinity can be
interpreted either as a net pumping flow or, in the frame moving with the sheet,  as a swimming velocity.

At leading order, i.e.~at order $\epsilon^2$, the flow induced by the sheet, and the rate of working by the sheet to induce
this flow, is the same as the Newtonian solution. The non-Newtonian contributions in both the induced net flow and the rate
of working appear at the next order, i.e.~$O(\epsilon^4)$. In both cases, the effect is linear in $n-1$ meaning that
shear-thinning fluids ($n<1$) and shear-thickening fluids ($n>1$) always induce opposite effects. The leading-order
non-Newtonian contribution to the flow created at infinity is non-zero only if the sheet deforms both in the direction normal
and tangential to the wave direction. The non-Newtonian contribution to the rate of working is found to always be negative in
the case of shear-thinning fluids, which are thus seen to systematically decrease the cost of transport independently of the
details of the wave kinematics. The maximum gain in dissipated energy is obtained for antiplectic waves where the direction
of the  propagating wave and that of the  effective cilia stroke are  opposite.

The sheet kinematics can be interpreted as a model for the locomotion of a flagellated microorganism. In this case, and
making the biologically-realistic assumption that the flagellum is inextensible, we find that the swimmer is more efficient
in a shear-thinning fluid but that  the non-Newtonian contribution to the swimming speed of the sheet is exactly equal to
zero. This result is consistent with recent experiments  \cite{Shen2011} showing that the swimming speed of a nematode waving
freely in a shear-thinning fluid is  same as that observed in a Newtonian fluid. In contrast, computations with swimmers
deforming their flagella  with increasing amplitude showed that the locomotion speed was enhanced by shear thinning. The
effect was attributed  to a viscosity gradient induced by the increased  beating flagellar amplitude, a result which could
potentially be tackled theoretically with an approach similar to ours adapted to their kinematics \cite{Smith2012}. In
contrast, recent experiments with a two-dimensional swimmer model showed a decrease of the swimming speed in a shear-thinning
elastic fluid \cite{dasgupta13}, a result not explained by our theoretical approach but perhaps due to the importance of
higher-order terms for large-amplitude swimming or due to non-negligible linear elastic effects.

Viewing the sheet as a model for the transport of non-Newtonian fluids by cilia arrays, we address three different types of
kinematics to understand the impact of shear-thinning fluids on the flow transport: kinematics dominated by normal
deformation, those dominated by tangential motion, and kinematics where normal and tangential beating were of the same order
of magnitude. In all three cases we consider four different values of the phase between the normal and tangential waving
motion, and the results are summarized in Table~\ref{tabletwo} where we used ``$+$" to denote instances when the
shear-thinning fluid improves on the performance (flow velocity or transport efficiency) and ``$-$'' otherwise. In all cases,
a shear-thinning fluid renders the flow transport more efficient, and in eight out of the twelve cases it also increases the
transport speed -- in particular for the biologically relevant case where the magnitudes of the normal and tangential
deformation are of the same order. A further numerical investigation in \S\ref{numerics} also identifies the regions in which
the non-Newtonian term is stronger than the Newtonian one.

\begin{table}[t]
  \centering
  \scalebox{0.9}{\begin{tabular}{c|c|c|c}
                      & \quad Normal + tangential \quad & \quad Normal mode \quad & \quad Tangential mode \quad  \\
                      & $a=b$ & $b\gg{a}$ & $a\gg{b}$\\
                                 &&& \\
\hline
           &&& \\
           $\phi=0$   & velocity: $+$ & velocity: $-$ & velocity: $-$ \\
                      &efficiency: $+$ & efficiency: $+$ &efficiency: $+$ \\
                      &&& \\
\hline
           &&& \\
           $\phi=-\pi/2,\pi/2$   & velocity: $+$ & velocity: $+$  & velocity: $-$  \\
                      &efficiency: $+$ & efficiency: $+$  &efficiency: $+$ \\
                      &&& \\
\hline
           &&& \\
           $\phi=\pi$   & velocity: $+$ & velocity: $+$ & velocity: $+$ \\
                      &efficiency: $+$ &efficiency: $+$&efficiency: $+$ \\
           &&& \\
  \end{tabular}}
  \caption{Effect of a shear-thinning viscosity on the mean transport velocity and  efficiency for  three different wave kinematics: normal plus tangential deformation, motion   dominated by the normal mode, and motion dominated by the tangential mode. Four different values of the phase between normal and tangential motion are considered.   The sign $+$ (resp.~$-$) indicates a situation where   a shear-thinning fluid improves on (resp.~hinders)  the velocity or efficiency; all signs would be  reversed in a shear-thickening fluid.}
  \label{tabletwo}
\end{table}

In this paper we have assumed that the fluid has no memory but instead its viscosity is an instantaneous function of the
shear rate -- a class of models known as generalized Newtonian fluids. Under that limit, we saw that  the  shear-thinning
property of mucus does facilitate its transport by cilia arrays.  The constitutive model we used in this work then serves as
an alternative model to the previously-studied two-fluid model and allows us to treat the mucus layer as a continuous fluid.
We still need, however, to estimate the contribution of the shear-thinning effects. Indeed, they do not play any role at
leading order but  only contribute at $\mathcal{O}(\epsilon^{4})$. Elastic stresses, in contrast, do contribute at  order
$\mathcal{O}(\epsilon^{2})$ and hinder the fluid transport by the prefactor \cite{Lauga2007,Shen2011}.
\def\De{{\rm De}}
\begin{equation}\label{elasticity}
\frac{1+\De^{2}\eta_{s}/\eta}{1+\De^{2}},
\end{equation}
where the Deborah number, $\De$, is defined as $\De=\lambda_{r}\omega$, with $\lambda_{r}$ being the relaxation time of the
fluid, and $\eta_{s}/\eta $ is the ratio of the solvent viscosity with the viscosity of the polymeric solution (typically
$\eta_{s}/\eta \ll 1$). For mucus we have $\lambda_{r}\approx 30$ s \cite{Gilboa1976}, and thus, using the typical frequency
in Table~\ref{tableone} for tracheobronchial cilia, we have $\De \approx 4\times 10^3\gg 1$. For a  solvent with the same
viscosity as water, we have   $\eta_{s}/\eta \approx 10^{-5} $ at zero shear rate (see Fig.~\ref{mucus}) and thus $\De^2
\eta_{s}/\eta\gg1$. Elastic effects lead therefore to a decrease of the transport speed by a factor of $\eta_{s}/\eta$.

To compare simply the elastic and shear-thinning contributions, consider a viscoelastic fluid being deformed by a
low-amplitude undulating surface with normal and tangential motion of the same order of magnitude. The  magnitude of the
shear-dependent contribution to the transport velocity scales as $\sim \epsilon^{4}|\mathcal{N}|\Cu^{2}$ while the elastic
contribution scales as $\sim \epsilon^{2}\eta_s/\eta $. With $|\mathcal{N}|\approx 0.4$,  $\lambda_{t}\approx2000$ s
(Fig.~\ref{mucus}), and  $\omega = 2\pi\times 20$ Hz (Table \ref{tableone}), the ratio of shear-dependent to elastic effects
varies as $\sim \epsilon^{2}|\mathcal{N}|\Cu^{2}/ (\eta_s/\eta )\sim 10^{15}\,\epsilon^2$. Despite the different scaling with
the motion amplitude $\epsilon$, and although our results are only strictly valid  in the mathematical limit $\epsilon \to
0$, we  expect that the shear-dependent viscosity will play the most important role on the ability of cilia to transport
mucus-like fluids. In addition, a similar scaling ratio can be computed to compare the non-Newtonian contribution derived in
our paper, $U^{(4)}_{NN}$, to the Newtonian speed at leading order, $U^{(2)}$, and we find $U^{(4)}_{NN}/U^{(2)}\sim
\epsilon^{4}|\mathcal{N}|\Cu^{2}/ \epsilon^2  \sim 10^{10}\,\epsilon^2 $, another potentially very large ratio\footnote{ We
emphasize again that the calculations presented in this paper are only valid in the mathematical limit $\epsilon \to 0$. The
order-of-magnitude estimates presented above are only formally true when $\epsilon$ is small and  are  used to point out how
easy it is for the shear-dependent terms to become predominant.}. It is therefore clear that  the rheological properties of
the surrounding fluid will play important roles in the fluid flow produced by cilia arrays
\cite{Sleigh1988,Sanderson1981,Sade1970}.

Beyond the assumption on the fluid rheology, the model proposed in this paper has been carried out under a number of
mathematical assumptions. The calculation was two-dimensional and as such does not address the flow in the direction
perpendicular to the wave propagation. For the sheet as a model for  flagellar locomotion, the finite-length of the flagellum
is expected to play an important role in non-linear fluids. And for cilia, the discrete nature of the cilia in a dense array
will also lead to nontrivial flow in the sublayer with contributions to both transport and energetics still unquantified. It
is  hoped that these limitations do not prevent the current results to still be  relevant  to both locomotion and transport
studies, for example  mucus transport in the respiratory track, but instead  suggest potentially interesting directions with
future work.

\section*{Acknowledgements}

R.~V\'elez acknowledges CONACyT-M\'exico and UC-MEXUS for their financial support during his postdoctoral stay at UCSD. We
also thank the support of the NSF (grant  CBET-0746285 to E.~L.).

\appendix
\section{Application to other generalized Newtonian fluids}
\label{appendix}
In this paper we have used a specific empirical model (Carreau fluid) to mathematically describe the relationship between  viscosity and shear rate. How applicable are our results to fluids described by other rheological laws, $\eta(\dot{\gamma})$? In the asymptotic limit  assumed this work, we need a fluid model able to describe the behavior of flows in the limit of low shear rates. We first require the viscosity, $\eta$, to be a smooth function of the shear rate, $\dot{\gamma}$,  leading a
nonzero value of $\eta$ in the limit $\dot{\gamma} \to 0$, denoted $\eta(0)$. This requirement rules out  the use of the power-law model, valid only in the limit of high shear rates  \cite{Bird1987}. Due to the $\epsilon \to - \epsilon$ symmetry,  the shear-dependent viscosity needs to be an even function of the shear rate. The only even function at order one in $\epsilon$ is the
absolute value function, which is not smooth (differentiable) at $\dot{\gamma}=0$. Hence, the viscosity necessarily has to be a function of the shear rate square.  As described in the text,  we  have
\begin{equation}\label{expansion.magnitude.appen}
|\dot{\gamma}|^{2}=\epsilon^{2}\frac{\Pi^{(1)}}{2}+\mathcal{O}(\epsilon^{3}).
\end{equation}
In the limit of small wave amplitudes, the viscosity  then takes the general form
\begin{equation}\label{appen.2}
\eta(|\dot{\gamma}|^{2})=\eta\left[\epsilon^{2}\frac{\Pi^{(1)}}{2}+\mathcal{O}(\epsilon^{3})\right]=\eta(0)+\epsilon^{2}\frac{\Pi^{(1)}}{2}\frac{d\eta}{d\dot{\gamma}^{2}}\bigg\vert_{0},
\end{equation}
using a Taylor expansion near zero. The shear-dependent viscosity will thus give an effect at order $\mathcal{O}(\epsilon^{2})$ at best, and thus at $\mathcal{O}(\epsilon^{3})$ in the flow field. If the  derivative in eq.~\eqref{appen.2} is zero then the effect will be at higher order. If instead the derivative is not defined, then the model is not appropriate to describe flow behavior at small shear rates. In the case of the Carreau model considered above, the derivative
${d\eta}/{d\dot{\gamma}^{2}}|_{0}$ takes a finite value, so we indeed obtain a nonzero effect at second order. There are other models that could potentially be used as well to fit rheological data. One of such model is that of a  Cross  fluid, given in a dimensionless form by
\begin{equation}\label{appen.cross}
{\eta}^{-1}=1+\left[(Cu|\dot{\gamma}|)^{2}\right]^{-\mathcal{N}}.
\end{equation}
In that case, it is first straightforward to show that only $\mathcal{N}\leq 0$ (shear-thinning fluid) leads to a finite
value of the viscosity at zero shear rate. Computing the derivative in eq.~\eqref{appen.cross} to perform the Taylor
expansion it is then easy to show that only the specific case $\mathcal{N}=-1$ leads to a finite value for
${d\eta}/{d\dot{\gamma}^{2}}|_{0}$. When $\mathcal{N}=-1$ the results obtained for a Cross fluid would thus be similar to the
results described in the paper; when  $\mathcal{N}>-1$ the Cross fluid is not well defined near the zero-shear-rate limit
while when  $\mathcal{N}<-1$ the derivative of the viscosity is zero, and therefore non-Newtonian effects on the fluid flow
would appear at a higher order. A similar analysis can be carried out for the Ellis fluid, given by,
\begin{equation}\label{appen.ellis}
{\eta}^{-1}=1+\left[{|\tau|^2}/{\tau^2_{(1/2)}}\right]^{\mathcal{N}}
\end{equation}
with similar results ($|\tau|$ is the magnitude of the stress tensor and $\tau_{(1/2)}$ is the shear at which $\eta=1/{2}$).
Consequently, for any generalized Newtonian fluid with a well-defined behavior in the zero-shear-rate limit, the asymptotic
results are   either  exactly the same as the ones in our paper, or they predict a non-Newtonian impact only occurring at
higher-order in the wave amplitude (and are thus identical to the Newtonian results up to  order $\epsilon^4$ at least).

\end{document}